\documentclass[%
superscriptaddress,
twocolumn,
nofootinbib,
 amsmath,amssymb,
 aps, physrev,
]{revtex4-2}

\usepackage{graphicx}% Include figure files
\usepackage{dcolumn}% Align table columns on decimal point
\usepackage{bm}% bold math
\usepackage{hyperref}% add hypertext 
\usepackage{orcidlink}
\usepackage{booktabs}

\newcommand{{\qsub}}{\bf{q}}
\newcommand{{\rsub}}{{\bf{r}}}
\newcommand{{\xlight}}{{\bf{n}}_{\rm{light}}}
\newcommand{{\xlens}}{{\bf{n}}_{\rm{mac}}}
\newcommand{{\data}}{\bf{D}}
\newcommand{{\datan}}{{\bf{d}}_{\rm{n}}}
\newcommand{{\datanprime}}{{{\bf{d}}_{\rm{n}}^{\prime}}}
\newcommand{{\dimg}}{{\bf{d}}_{\mathcal{I}}}
\newcommand{{\dimgprime}}{{\bf{d}}_{\mathcal{I}}^{\prime}}
\newcommand{{\dptsrc}}{{\bf{d}}_{\rm{ptsrc}}}
\newcommand{{\dptsrcprime}}{{\bf{d}}_{\rm{ptsrc}}^{\prime}}
\newcommand{{\dfr}}{{\bf{d}}_{\rm{fr}}}
\newcommand{{\dfrprime}}{{\bf{d}}_{\rm{fr}}^{\prime}}
\newcommand{{\zlens}}{{z_{\rm{d}}}}
\newcommand{{\zsrc}}{{z_{\rm{s}}}}
\newcommand{{\nbody}}{{$N$-body~}}
\newcommand{{\msun}}{{\rm{M}_{\odot}}}

\begin{document}

\preprint{APS/123-QED}

\title{
The free-streaming length of dark matter from JWST observations of 28 strong gravitational lenses} 
\date{\today}

    \author{D.~Gilman$^\star$ \orcidlink{0000-0002-5116-7287}}
    \thanks{Brinson Prize Fellow}
    \email{gilmanda@uchicago.edu}
	\affiliation{Department of Astronomy \& Astrophysics, University of Chicago, Chicago, IL 60637, USA}
    
    \author{A.~M.~Nierenberg$^\star$ \orcidlink{0000-0001-6809-2536}}
    \email{anierenberg@ucmerced.edu}
	\affiliation{University of California, Merced, 5200 N Lake Road, Merced, CA 95341, USA}

    \author{T.~Treu\orcidlink{0000-0002-8460-0390}}
	\affiliation{Department of Physics and Astronomy, University of California, Los Angeles, CA,  90095, USA}
    
    \author{K.~N.~Abazajian\orcidlink{0000-0001-9919-6362}}
    \affiliation{Department of Physics and Astronomy, University of California, Irvine, CA 92697-4575, USA}
    
    \author{T.~Anguita\orcidlink{0000-0003-0930-5815}}
    \affiliation{Instituto de Astrofisica, Departamento de Fisica y Astronomia, Universidad Andres Bello, Chile}

    \author{V.~N.~Bennert\orcidlink{0000-0003-2064-0518}}
    \affiliation{Physics Department, California Polytechnic State University, San Luis Obispo, CA 93407, USA }

    \author{A.~J.~Benson\orcidlink{0000-0001-5501-6008}}
    \affiliation{Carnegie Institution for Science, Pasadena CA 91101, USA }

    \author{S.~Birrer\orcidlink{0000-0003-3195-5507}}
    \affiliation{Department of Physics and Astronomy, Stony Brook University, Stony Brook, NY 11794, USA}
    
    \author{S.~G.~Djorgovski\orcidlink{0000-0002-0603-3087}}
    \affiliation{California Institute of Technology, Pasadena CA 91125, USA }

    \author{X.~Du\orcidlink{0000-0003-0728-2533}}
    \affiliation{Department of Physics and Astronomy, University of California, Los Angeles, CA,  90095, USA}
    
    \author{C.~Gannon\orcidlink{0009-0009-0443-3181}}
    \affiliation{University of California, Merced, 5200 N Lake Road, Merced, CA 95341, USA}
    
    \author{S.~F.~Hoenig\orcidlink{0000-0002-6353-1111}}
    \affiliation{School of Physics and Astronomy, University of Southampton, Southampton SO17 1BJ, United Kingdom }

    \author{R.~E.~Keeley\orcidlink{0000-0002-0862-8789}}
    \affiliation{University of California, Merced, 5200 N Lake Road, Merced, CA 95341, USA}
    
    \author{A.~Kusenko\orcidlink{0000-0002-8619-1260}}
    \affiliation{Department of Physics and Astronomy, University of California, Los Angeles, CA,  90095, USA}
    \affiliation{Kavli Institute for the Physics and Mathematics of the Universe (WPI), UTIAS, The University of Tokyo, Kashiwa, Chiba 277-8583, Japan}

    \author{H.~R.~Larsson\orcidlink{0000-0002-9417-1518}}
    \affiliation{University of California, Merced, 5200 N Lake Road, Merced, CA 95341, USA}

    \author{M.~Malkan\orcidlink{0000-0001-6919-1237}}
    \affiliation{Department of Physics and Astronomy, University of California, Los Angeles, CA,  90095, USA}
    
    \author{T.~Morishita\orcidlink{0000-0002-8512-1404}}
    \affiliation{IPAC, California Institute of Technology, MC 314-6, 1200 E. California Boulevard, Pasadena, CA 91125, USA}
    
    \author{V.~Motta\orcidlink{0000-0003-4446-7465}}
    \affiliation{Instituto de F\'{\i}sica y Astronom\'{\i}a, Universidad de Valpara\'{\i}so, Avda. Gran Breta\~na 1111, Valpara\'{\i}so, Chile}

    \author{L.~A.~Moustakas\orcidlink{0000-0003-3030-2360}}
    \affiliation{Jet Propulsion Laboratory, California Institute of Technology, 4800 Oak Grove Dr, Pasadena, CA 91109}

    \author{P.~Mozumdar\orcidlink{0000-0002-8593-7243}}
    \affiliation{Department of Physics and Astronomy, University of California, Los Angeles, CA,  90095, USA}
    \affiliation{Department of Physics and Astronomy, University of California, Davis, 1 Shields Ave., Davis, CA 95616, USA}
    
    \author{H.~Paugnat\orcidlink{0000-0002-2603-6031}}
    \affiliation{Department of Physics and Astronomy, University of California, Los Angeles, CA,  90095, USA}
    
    \author{W.~Sheu\orcidlink{0000-0003-1889-0227}}
    \affiliation{Department of Physics and Astronomy, University of California, Los Angeles, CA,  90095, USA}
    
    \author{D.~Sluse\orcidlink{0000-0001-6116-2095}}
    \affiliation{STAR Institute, University of Li{\`e}ge, Quartier Agora, All\'ee du six Ao\^ut 19c, 4000 Li\`ege, Belgium}
    
    \author{D.~Stern\orcidlink{0000-0003-2686-9241}}
    \affiliation{Jet Propulsion Laboratory, California Institute of Technology, 4800 Oak Grove Dr, Pasadena, CA 91109}

   \author{M.~Stiavelli\orcidlink{0000-0002-8512-1404}}
    \affiliation{Space Telescope Science Institute, 3700 San Martin Drive, Baltimore, MD 21218, USA}

    \author{D.~Williams\orcidlink{0000-0002-8386-0051}}
    \affiliation{Department of Physics and Astronomy, University of California, Los Angeles, CA,  90095, USA}

    \author{K.~C.~Wong\orcidlink{0000-0002-8459-7793}}
    \affiliation{Research Center for the Early Universe, Graduate School of Science, The University of Tokyo, 7-3-1 Hongo, Bunkyo-ku, Tokyo 113-0033, Japan}
             
\begin{abstract}
The formation of gravitationally bound overdensities of dark matter (DM), or \textit{halos}, is a generic prediction of theories with particle DM. We present a measurement of halo properties in 28 quadruple image strong lens systems recently observed by JWST, and use these observations to constrain the free-streaming length, $\lambda_{\rm{FS}}$, of DM, a quantity that depends on the DM particle mass and formation mechanism. We improve on previous lensing analyses by simultaneously reconstructing extended lensed arcs with image positions and relative magnifications, enhancing sensitivity to perturbations by halos. Our analysis rules out deviations from the predictions of cold dark matter (CDM) on scales above $10^{7.2} M_{\odot}$ and $10^{7.4} M_{\odot}$ for subhalo abundance predicted by cosmological $N$-body simulations and semi-analytic models, respectively. These bounds correspond to upper limits $\lambda_{\rm{FS}}<6.0 \ \rm{kpc}$ and $\lambda_{\rm{FS}}<7.0 \ \rm{kpc}$, and lower limits on the mass of a spin--1/2 thermal relic DM particle $m_{\rm{therm}}>7.4 \ \rm{keV}$ and $m_{\rm{therm}}>6.5 \ \rm{keV}$. Conversely, assuming a negligible free-streaming length, as predicted by CDM, we measure a projected mass in subhalos around elliptical galaxies $1.7_{-1.2}^{+2.6} \times 10^7 \ \mathrm{M}_{\odot} \ \rm{kpc^{-2}}$ at $95 \%$ confidence. These inferences confirm key predictions of the CDM paradigm. 

\end{abstract}

\maketitle

\def\thefootnote{$\star$}\footnotetext{These authors contributed equally to this work}\def\thefootnote{\arabic{footnote}}

%\def\thefootnote{$\star$}
%\footnotetext{These authors contributed equally to this work}
%\def\thefootnote{\arabic{footnote}}

%\section{\label{sec:level1}Introduction}

%{\bf{{Introduction}}} 
\textit{Introduction}---The physical properties of dark matter, such as its particle mass, formation mechanism, or possible self-interactions, remain one of the most important unsolved questions in cosmology and modern physics. Cosmic gravitational probes of dark matter provide direct avenues to investigate these open questions \citep{BuckleyPeter18,Boddy++22,DrlicaWagner22}. Cosmic probes, in most cases, do not aim to directly detect dark matter particles. Instead, they exploit the connection between the particle physics of dark matter and the properties of dark matter \textit{halos}, self-gravitating overdensities of dark matter that permeate the cosmos. 

This Letter presents an inference on a key scale related to dark matter: the free-streaming length, $\lambda_{\rm{FS}}$. Free-streaming refers to the process by which dark matter particles escape from density fluctuations initialized in the early Universe \citep{Bond++83,AvilaReese++01,Bode++01}. At later times, free-streaming manifests itself as a suppression of the overall abundance and concentrations of halos below a characteristic scale \citep{Schneider++13,Bose++16,Ludlow++16}. This scale depends on the velocity distribution of the dark matter particle at early times, which means inferences of halo abundance and structure can be recast as inferences on the dark matter particle mass and production mechanism. Dark matter models in which the free-streaming length becomes $\mathcal{O}\left(1 \ \rm{kpc}\right)$ cause a deviation from the predictions of cold dark matter (CDM) on mass scales comparable to dwarf galaxies, and are categorized as warm dark matter (WDM). 

Strong gravitational lensing enables the detection of dark matter halos at cosmological distances, irrespective of whether halos contain detectable gas or stars. In a strong gravitational lens, a foreground deflector, typically a massive elliptical galaxy \citep{Auger++10}, produces multiple magnified images of a background source. Dark matter halos, both in the plane of the lens and along the entire line of sight, impart perturbations to the image positions and magnifications on angular scales much smaller than the typical size of the main deflector. Characterizing these perturbations enables inferences of halo properties directly through gravity, and is emerging as a leading cosmic probe of dark matter  \citep{Dalal++02,Vegetti++12,Hezaveh++16,Birrer++17,Despali++18,Gilman++19,Gilman++20,Hsueh++20,Gilman++22,Gilman++23,Powell++25}.

This analysis considers a particular class of strong lens system in which a background quasar becomes quadruply imaged. The magnifications of lensed images depend on second derivatives of the gravitational potential projected onto the plane of the lens, and the sensitivity to small-scale structure depends on the angular size of the source relative to the deflection angle produced by a perturber \citep{Dobler++06}. For compact sources, such as a quasar, low-mass halos along the light travel path can significantly (de)magnify the images \citep{MaoSchneider98}. However, a very compact source can be substantially (de)magnified by stars in the main deflector galaxy, a phenomenon known as optically thick microlensing \citep{Wambsganss_microlensing}. Studies of dark matter substructure from lensed quasars must therefore measure emission from an extended area around the quasar to avoid microlensing contamination. 

This Letter reports a new measurement of the free-streaming length of dark matter enabled by two major advances in the field of strong lensing investigations of dark matter substructure. First, we measure relative magnifications (flux ratios) using the JWST Mid-Infrared Instrument (MIRI) in 26 lensed quasars, doubling the sample size of lens systems suitable for a dark matter analysis, relative to previous work. The MIRI observations detect mid-infrared emission from the warm dust region surrounding the background quasar, which extends over 1--10 pc---large enough to avoid optically thick microlensing \citep{Sluse++13, burtscher_diversity_2013, leftley_parsec-scale_2019, honig_redefining_2019}, yet compact enough to experience perturbations by dark substructures as small as $10^6$ M$_{\odot}$ \citep{Nierenberg++24}. The second significant improvement implemented in this analysis involves the joint modeling of image positions and flux ratios with lensed extended emission from the quasar host galaxy (hereafter lensed arcs). Simultaneous modeling of these data isolates small-scale perturbations by halos to the flux ratios from uncertainties associated with the mass profile of the main deflector. Companion papers describing the observations and flux ratio measurements \citep{Keeley++25}, and the lens modeling and inference techniques \citep{Gilman++25} (hereafter G25) give additional details on the methods and results presented in this Letter. 

\textit{Free-streaming in warm dark matter models}---Free-streaming of a dark matter particle that comprises $100 \%$ of the dark matter manifests itself as a cutoff in the linear matter power spectrum. Following convention \citep{Viel++05}, we parametrize the WDM power spectrum in terms of a transfer function $T\left(k\right)$
\begin{equation}
    \label{eqn:transferfunc}
    \frac{P_{\rm{wdm}}\left(k\right)}{P_{\rm{cdm}}\left(k\right)} \equiv T^2\left(k\right) = \left(1+\left( \alpha k \right)^{2\nu}\right)^{-10/\nu}.
\end{equation}
The effects of WDM on halo properties become significant when the WDM power spectrum deviates significantly from the CDM prediction. This occurs on scales smaller than the half-mode wavenumber $k_{\rm{hm}}$, defined via $T\left(k_{\rm{hm}}\right)=1/2$, which corresponds to the half-mode mass
\begin{equation}
\label{eqn:mhm}
m_{\rm{hm}} = \left(4 \pi /3 \right) \Omega_{\rm m} \rho_{\rm{crit}} \left(\pi / k_{\rm{hm}}\right)^3,
\end{equation}
where $\Omega_m$ and $\rho_{\rm{crit}}$ are the matter density and critical density of the Universe. The free-streaming length itself, $\lambda_{\rm{FS}}$, corresponds approximately to the comoving distance traversed by dark matter particles before density fluctuations start growing appreciably, which occurs around the epoch of matter-radiation equality \citep{Kolb90}. In practice, it is more common to define an effective free-streaming scale from Equation \ref{eqn:transferfunc} as $\lambda_{\rm{FS}}\equiv \alpha$. In our adopted cosmology \citep{Planck++20}, this gives 
\begin{equation}
\label{eqn:lambdaFS}
\lambda_{\rm{FS}}=5.2\left(\frac{m_{\rm{hm}}}{10^7 M_{\odot}}\right)^{1/3} \ \rm{kpc}.
\end{equation}

The shape of the transfer function depends on the mass and formation mechanism of the WDM particle. We consider the case of thermal relic particles of mass $m_{\rm{therm}}$, which we can relate to the half-mode mass
\begin{equation}
    \label{eqn:mhmmtherm}
    m_{\rm{hm}} = M_0 \left(\frac{m_{\rm{therm}}}{3 \,\rm{keV}}\right)^{\zeta} \mathrm{M}_{\odot}.
\end{equation}
In our adopted cosmology, for spin--1/2 particles we have $\left(M_0,\zeta,\nu\right) = \left(4.0 \times 10^8 M_{\odot},-3.564,1.049\right)$, while for spin-3/2 particles $\left(M_0,\zeta,\nu\right) = \left(2.1 \times 10^8 M_{\odot},-3.585,1.025\right)$ \citep{Vogel++23}. While this relation is calibrated for thermal relic WDM, it proves a useful proxy for many other WDM models, such as sterile neutrinos, that possess a similarly shaped transfer function but typically have $m_\mathrm{hm}$ and $\lambda_\mathrm{FS}(\alpha)$ correspond to a different particle mass than $m_\mathrm{therm}$  \citep{Kusenko:2006rh,Petraki:2007gq,Petraki:2008ef,Kusenko:2009up,Abazajian:2019ejt,Zelko++22,Stucker++22}. 

The effects of free-streaming on halo abundance and density profiles are borne out through high resolution cosmological simulations of nonlinear structure formation through $N$-body simulations and semi-analytic models. These simulations relate the halo mass function in CDM, $dN_{\rm{cdm}}/dm$, to $m_{\rm{hm}}$ through a fitting function
\begin{equation}
\label{eqn:mfunc}
\frac{dN_{\rm{wdm}}}{dm}\left(m,z\right) = \frac{dN_{\rm{cdm}}}{dm}\left(m,z\right)\left[1+\left(a\frac{m_{\rm{hm}}}{m}\right)^b\right]^c,
\end{equation}
and give a similar parameterization of the concentration-mass relation
\begin{equation}
\label{eqn:mcrel}
c_{\rm{wdm}}\left(m,z\right) = c_{\rm{cdm}}\left(m,z\right) \left[1+\left(a^{\prime}\frac{m_{\rm{hm}}}{m}\right)^{b^{\prime}}\right]^{c^{\prime}}.
\end{equation}
We use fitting functions for the mass function and concentration-mass relation presented by \citet{Lovell++20} and \citet{Bose++16}, respectively, which give $\left(a,b,c\right)=\left(2.1,0.8,-1.0\right)$ and $\left(a^{\prime},b^{\prime},c^{\prime}\right)=\left(60,1.0,-0.17\right)$. For the CDM concentration-mass relation, we use the model presented by \citet{DiemerJoyce19}. We model halos as Navarro-Frenk-White \citep[][hereafter NFW]{Navarro++97} profiles, and define halo mass in terms of the virial radius $r_{\rm{vir}}=r_{200}$ that encloses a mean density of $200\rho_{\rm{crit}}\left(z\right)$. 

We use Equations \ref{eqn:mfunc} and \ref{eqn:mcrel} to model dark matter substructure in the lens systems. For the line of sight population, we use Equation \ref{eqn:mfunc} with the model presented by \citet{ST01} in place of $dN_{\rm{cdm}}/dm$, plus an enhancement of structure near the main deflector from clustering effects predicted by hierarchical structure formation \citep{Gilman++19,Lazar++21}, and an overall rescaling factor, $\delta_{\rm{LOS}}$, to account for variation in the number of line-of-sight halos from uncertainties in cosmological parameters and the halo mass function model. 

For subhalos, or halos gravitationally bound to the main deflector that orbit within the tidal field of a host, we use a mass function of the form 
\begin{eqnarray}
\label{eqn:shmf}
\frac{d^2N}{dm dA} = \frac{\Sigma_{\rm{sub}}}{m_0}\left(m/m_0\right)^{\alpha}\mathcal{F}\left(m_{\rm{host}},z_{\rm{d}}\right) \\
\nonumber\times\left[1+\left(a\frac{m_{\rm{hm}}}{m}\right)^b\right]^c
\end{eqnarray}
with a pivot scale $m_0=10^8 M_{\odot}$. The parameters $\Sigma_{\rm{sub}}$ (units kpc$^{-2}$) and $\alpha$ set the normalization and logarithmic slope. The term $\mathcal{F}\left(m_{\rm{host}},z\right) = \left(m_{\rm{host}}/10^{13} \mathrm{M}_{\odot}\right)^{k_1}\left(z+0.5\right)^{k_2}$, with $k_1=0.55$ and $k_2=-0.37$ \citep{Gannon++25}, accounts for the scaling of the projected number density of subhalos with the host halo mass $m_{\rm{host}}$ and redshift $z_{\rm{d}}$. By including this term, $\Sigma_{\rm{sub}}$ and $\alpha$ become hierarchical parameters of the model, meaning they are common parameters to each lens in the sample. Given $\Sigma_{\rm{sub}}$, $\alpha$, and $m_{\rm{host}}$, we generate a population of subhalos with masses and concentrations defined at infall. We then use the tidal evolution model presented by \citet{Du++25} to predict the density profile of each subhalo, which we model as a tidally truncated NFW profile \citep{Baltz++09,Du++24}. 

We render halos in the mass range $10^6 - 10^{10.7} M_{\odot}$. Halos less massive than $10^6 M_{\odot}$ are too small to affect the data, while halos more massive than $10^{10.7} M_{\odot}$ host detectable galaxies, and we include these objects explicitly the lens model at their observed positions. Section IV in G25 provides further discussion regarding the modeling of dark matter substructure used in our simulations. 

\textit{Observations}---We compiled several datasets to perform the dark matter inference. Measurements of the relative point source brightnesses, or flux ratios, provide small scale (milliarcsecond) probes of the second derivative of the gravitational potential, while imaging of the extended lensed arcs and image positions provide an arcsecond scale constraint, primarily of the first derivative of the gravitational potential. 

The JWST lensed quasar dark matter survey (GO-2046) consists of observations of quadruply imaged quasars with MIRI targeting emission from the warm dust region surrounding the background quasar. This region is observed at low redshifts to be $\sim$1--10 pc in size with little dependence on the quasar luminosity \citep{burtscher_diversity_2013,leftley_parsec-scale_2019,honig_redefining_2019}. The warm dust region is compact enough to experience millilensing perturbations by halos, but extended enough (by $\gtrsim 1$ milliarcsecond at a typical source redshift $z \sim 2$) to remove contamination from stellar microlensing. The spectral energy distribution of warm dust emission peaks at observed (rest frame) wavelengths of $\sim$20 $\mu$m ($\sim$10 $\mu$m). Given typical lensed image separations of a few tenths of arcseconds, JWST MIRI is the first observatory and instrument capable of reaching the necessary wavelength, with sufficient sensitivity and spatial resolution, to make these measurements. \citet{Nierenberg++24} and \citet{Keeley++25} describe the lens selection criteria for the survey, and detail the spectral energy distribution fitting procedure used to separate warm dust emission from potentially microlensed emission emanating from the more compact hot dust torus and quasar accretion disk.

We supplement the 26 lenses observed by JWST with 2 systems that were not observed through the MIRI program. For these systems, we use nuclear narrow-line flux ratios using emission lines in the rest-frame optical (observed frame near-IR) with Keck OSIRIS with Adaptive Optics \citep{Nierenberg++14} and the HST WFC3 grism \citep[][]{Nierenberg++20}. The nuclear narrow-line region of quasars has typical physical sizes of $\sim$10-100 pc \citep{MoustakasMetcalf03,Muller-Sanchez++11, Nierenberg++14, Nierenberg++17} making it also insensitive to microlensing. 

We detect extended lensed arcs in 24/28 systems, and model the imaging data with the highest signal to noise ratio: HST F814W and F160W observations (10 systems, HST-GO-15320 and HST-GO-15652 PI:Treu), JWST NIRCam F115W (3 systems, JWST GTO-1198 PI Stiavelli) and JWST MIRI F560W (11 systems, JWST GO 2046 PI:Nierenberg), as summarized by G25 in Table II.  

\textit{Analysis Method}---Our goal is to infer the joint distribution of the dark matter model parameters ${\bf{q}}=\left(\Sigma_{\rm{sub}},\rm{m_{\rm{hm}}},\alpha,\delta_{\rm{LOS}}\right)$ given the data, ${\bf{D}}=\left({\bf{d}}_1,{\bf{d}}_2,..., \datan \right)$, where for the $n$th lens the data vector $\datan$ consists of the image positions, flux ratios, and the surface brightness of the lensed arcs. The hyper-parameters $\qsub$ specify statistical properties of the halo population, such as the average concentration of halos, or the amplitude of the (sub)halo mass function. The data depend on $\qsub$ through particular configurations, or \textit{realizations} $\rsub$, of halos. To evaluate the likelihood of the data given the model, we must therefore marginalize over many possible realizations
\begin{equation}
\label{eqn:likelihood}
\mathcal{L}\left(\datan | \qsub\right) = \int p\left(\datan | \rsub, \boldsymbol{\xi}\right)p\left(\rsub | \qsub \right) p\left(\boldsymbol{\xi}\right) d \rsub d \boldsymbol{\xi}\, ,
\end{equation}
where $\boldsymbol{\xi}$ represents nuisance parameters that describe the lens mass profile, the lens light profile, the lensed quasar host galaxy, the size of warm dust or nuclear narrow-line region from which we measure the flux ratios, the host halo mass, and the positions and masses of globular clusters near lensed images. 

The likelihood function involves an integral over all possible configurations of the main deflector mass profile and the positions, masses, and density profiles of individual (sub)halos. Our model for the main deflector mass profile is based on the observed structure of nearby massive elliptical galaxies \citep{Bender++88,Gavazzi_galaxy_density,Auger++10,Oh++24}, and includes azimuthal perturbations around isodensity contours (in projection) through an elliptical multipole expansion \citep{Paugnat++25}. In previous analyses, which used observations of image positions and flux ratios only, the lensing signal of low mass halos could be absorbed by these azimuthal features, and other transformations of the main deflector mass profile \citep[e.g.][]{Trotter++2000,Congdon++05,Cohen++24,Gilman++24}. As we discuss further in Paper IV (see in particular Section VI A), the extended arcs constrain the lens model across a larger area of the image plane, breaking some degeneracies between small scale structure and the mass profile of the main deflector. As dictated by Equation \ref{eqn:likelihood}, we marginalize over any remaining degeneracies when computing the likelihood function.

Given the likelihood function for individual lens systems from Equation \ref{eqn:likelihood}, we compute the joint posterior distribution for the population
\begin{equation}
p\left(\qsub | \bf{D}\right) \propto \pi\left(\qsub\right) \prod_{n=1}^{28} \mathcal{L}\left(\datan | \qsub\right)\, ,
\end{equation}
where $\pi\left(\qsub\right)$ is the prior probability. We assume a uniform prior on $\delta_{\rm{LOS}}$ within 0.9--1.1, allowing for $\pm 10 \%$ variations in the number of line-of-sight halos relative to the predictions of the Sheth-Tormen halo mass function model \citep{ST01}. We sample $\alpha$ between -1.95 and -1.85, which spans the range of theoretical uncertainties from simulations of non-linear structure formation \citep{Springel++08,Fiacconi++16,Benson20,Nadler++23}. We sample $\log_{10}m_{\rm{hm}}/M_{\odot}$ between 4--10. While $m_{\rm{hm}}=10^4 M_{\odot}$ is not formally consistent with CDM, these models are indistinguishable from CDM, given our data. Finally, we sample $\log_{10}\Sigma_{\rm{sub}} / \rm{kpc^{-2}}$ between -2.2 and 0.2. We sample the background source size for the lenses observed by JWST MIRI from a uniform prior between 1-10 pc, and the source size for the systems with narrow-line flux ratios from a uniform prior between 20-80 pc.

We use a forward modeling approach validated on simulated datasets \citep{Gilman++19,Gilman++21,Gilman++24} to evaluate Equation \ref{eqn:likelihood}. For each realization $\rsub$, we construct a lens model, compute model-predicted datasets $\datanprime\left(\rsub, \boldsymbol{\xi}\right)$ using multi-plane ray tracing, and compare these synthetic datasets with the observations to calculate the probability of the corresponding hyper-parameters $\qsub$. We evaluate Equation \ref{eqn:likelihood} by generating 0.5--22 million realizations per lens (176 million realizations total). We perform gravitational lensing calculations with {\tt{lenstronomy}}\footnote{\url{https://github.com/lenstronomy/lenstronomy}} \citep{Birrer++18,Birrer++21}, and generate populations of dark matter substructure using {\tt{pyHalo}}\footnote{\url{https://github.com/dangilman/pyHalo}}\citep{Gilman++20}. 

\begin{figure}
    \includegraphics[trim=1cm 1cm 1cm
    0.5cm,width=0.48\textwidth]{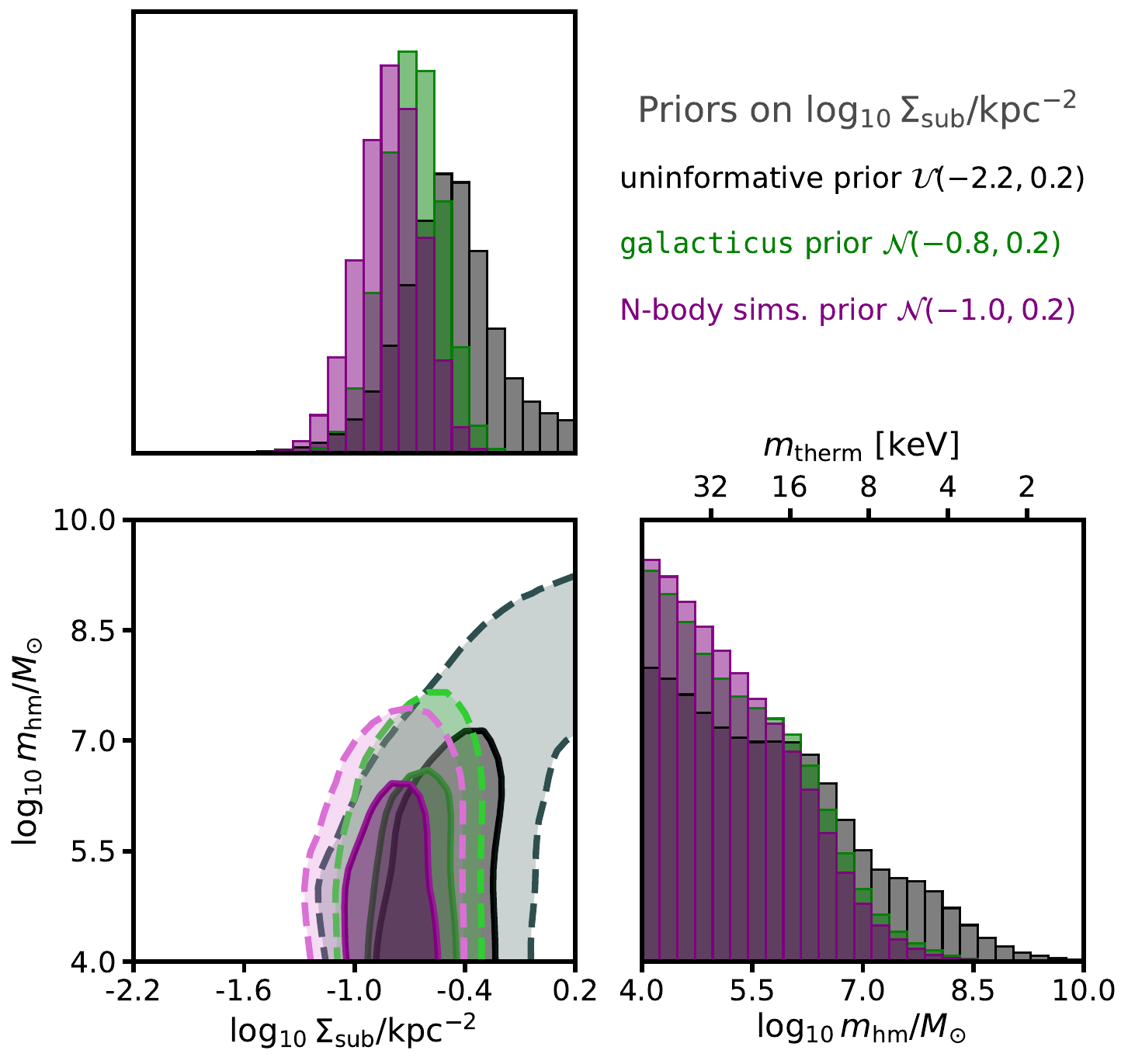}
    \caption{\label{fig:mhminference} The joint posterior probability distribution for the normalization of the projected subhalo mass function ($\Sigma_{\rm{sub}}$, Equation \ref{eqn:shmf}) and half-mode mass ($m_{\rm{hm}}$, Equation \ref{eqn:mhm}), jointly inferred from 28 lenses. The upper x-axis on the marginal distribution for $m_{\rm{hm}}$ shows the corresponding spin-1/2 thermal relic particle mass. We show three different priors on $\Sigma_{\rm{sub}}$. One-side $95\%$ exclusion limits, and 10:1 Bayes factor constraints taken with respect to the peak of the marginal distribution, are summarized in Table \ref{tab:constraints}. Contours show $68\%$ and $95\%$ confidence regions.}
\end{figure}

\begin{table*}
    \setlength{\tabcolsep}{12pt}
    \caption{\label{tab:constraints} Lower bounds on the thermal relic particle mass $m_{\rm{therm}}$ derived from constraints on the half-mode mass $m_{\rm{hm}}$ under different priors on $\log_{10}\Sigma_{\rm{sub}}/\rm{kpc}^{-2}$. We quote both the one-sided $95 \%$ exclusion limits, and the value of $m_{\rm{mtherm}}$ disfavored at 10:1 odds (or a 10:1 Bayes factor).}
    \begin{tabular}{llcccc}
        \hline
        & & \multicolumn{2}{c}{Spin-1/2} & \multicolumn{2}{c}{Spin-3/2} \\
        \cmidrule(lr){3-4} \cmidrule(lr){5-6}
        Prior on $\log_{10}\Sigma_{\rm{sub}}/\rm{kpc}^{-2}$ & Distribution 
            & 95\% excl. & 10:1 Bayes 
            & 95\% excl. & 10:1 Bayes \\
        \hline
        Uninformative   & $\mathcal{U}(-2.2,\,0.2)$   & $>4.7\,\rm{keV}$ & $>3.0\,\rm{keV}$ & $>3.9\,\rm{keV}$ & $>2.5\,\rm{keV}$ \\[4pt]
        \textsc{galacticus} SAM & $\mathcal{N}(-0.8,\,0.2)$   & $>9.6\,\rm{keV}$ & $>6.5\,\rm{keV}$ & $>8.0\,\rm{keV}$ & $>5.4\,\rm{keV}$ \\[4pt]
        $N$-body simulations    & $\mathcal{N}(-1.0,\,0.2)$   & $>10.2\,\rm{keV}$ & $>7.4\,\rm{keV}$ & $>8.5\,\rm{keV}$ & $>6.2\,\rm{keV}$ \\
        \hline
    \end{tabular}
\end{table*}
    
\textit{Results}---Figure \ref{fig:mhminference} shows the joint posterior distribution $p\left(\Sigma_{\rm{sub}}, m_{\rm{hm}}|\data\right)$, including the normalization of the subhalo mass function $\Sigma_{\rm{sub}}$ and the half-mode mass $m_{\rm{hm}}$. We marginalize over $\delta_{\rm{LOS}}$ and $\alpha$, the rescaling of the line-of-sight halo mass function and the logarithmic slope of the subhalo mass function, because these parameters are unconstrained within the range of the prior. The black distribution shows the inference with log-uniform priors on each hyper-parameter. For $m_{\rm{hm}}<10^{5} M_{\odot}$, the halo population is effectively indistinguishable from CDM with our data. For warmer models $\Sigma_{\rm{sub}}$ and $m_{\rm{hm}}$ are correlated. Warmer models with additional massive subhalos can mimic the lensing signal of colder models with fewer massive subhalos. 

The semi-analytic model {\tt{galacticus}} \citep{Benson12} and the Symphony $N$-body simulations \citep{Nadler++23} predict the amplitude of the bound subhalo mass function, meaning the amplitude of the subhalo mass function after accounting for tidal stripping. These predictions effectively describe the subhalo mass function amplitude on scales $m \gg m_{\rm{hm}}$, and correspond to $\Sigma_{\rm{sub}}\sim 0.15 \ \rm{kpc^{-2}}$ and $\Sigma_{\rm{sub}} \sim 0.1 \ \rm{kpc^{-2}}$, respectively \citep{Gannon++25,Gilman++25}. The green and magenta probability densities shown in Figure~\ref{fig:mhminference} show the inference after incorporating Gaussian priors on $\Sigma_{\rm{sub}}$ centered on these predictions, with a width of 0.2 dex, motivated by the differences between the predictions. The priors break covariance between $\Sigma_{\rm{sub}}$ and $m_{\rm{hm}}$. We infer (Bayes factor penalty of 10:1) upper bounds on $m_{\rm{hm}}$ of $10^{7.4} M_{\odot}$ and $10^{7.2} M_{\odot}$, for subhalo abundance predicted by {\tt{galacticus}} and Symphony, respectively. These constraints correspond to upper limits on the effective free-streaming scale (Equation \ref{eqn:lambdaFS}) $\lambda_{\rm{FS}}<7.0 \ \rm{kpc}$ and $\lambda_{\rm{FS}}<6.0 \ \rm{kpc}$, respectively. As summarized in Table \ref{tab:constraints}, these constraints translate to lower bounds on the mass of a spin--1/2 thermal relic particle of $m_{\rm{therm}}>6.5 \ \rm{keV}$ and $m_{\rm{therm}}>7.4 \ \rm{keV}$. For spin--3/2 particles, the bounds correspond to $m_{\rm{therm}}>5.4 \ \rm{keV}$ and $m_{\rm{therm}}>6.2 \ \rm{keV}$, respectively.

\begin{figure}
		\centering
		\includegraphics[trim=0.0cm 0.2cm 0.0cm
		0cm,width=0.48\textwidth]{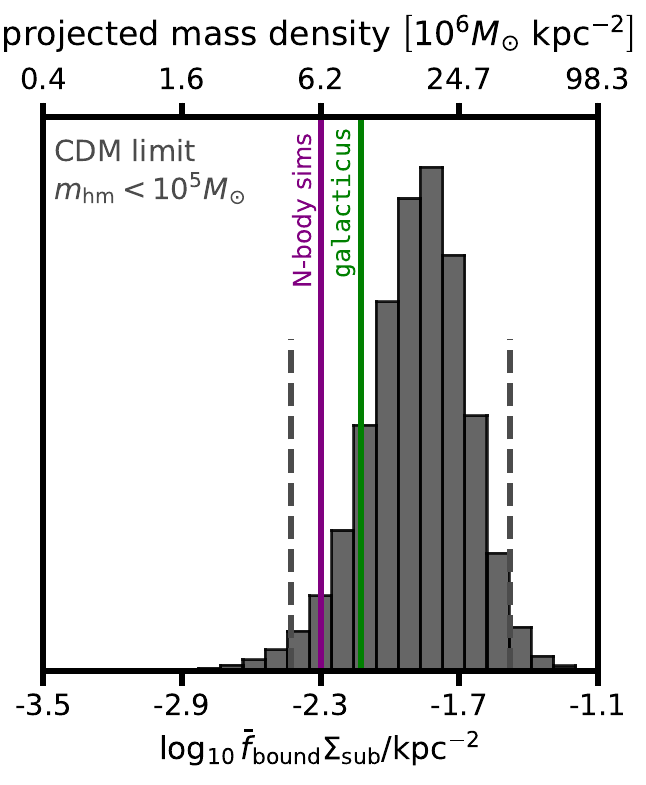}
		\caption{\label{fig:sigmasubinference} The posterior distribution of $\Sigma_{\rm{sub}}$ obtained from taking the CDM limit through a prior on the half-mode mass $\pi\left(\log_{10} m_{\rm{hm}} / \mathrm{M}_{\odot}\right) = \mathcal{N}\left(4.0, 0.5\right)$. The color scheme is the same as the left panel of Figure \ref{fig:mhminference}. Vertical dashed lines represent $95\%$ confidence intervals. Predictions from the semi-analytic model $\tt{galacticus}$ and $N$-body simulations are marked with vertical lines. The upper x-axis shows the corresponding projected mass density in subhalos with infall masses in the range $10^6 < m/M_{\odot}<10^{10.7}$.}
	\end{figure}

Figure~\ref{fig:sigmasubinference} shows the inference on subhalo abundance after assuming a prior on $\log_{10}m_{\rm{hm}}/M_{\odot} \sim \mathcal{N}\left(4.0, 0.5\right)$, which considers models in which $m_{\rm{hm}}$ becomes too small to affect the data. The lower x-axis shows constraints on the bound mass function amplitude $\bar{f}_{\rm{bound}}\Sigma_{\rm{sub}}$, where the mean bound mass fraction $\bar{f}_{\rm{bound}}=0.05$ is approximately independent of subhalo infall mass in CDM \citep{Du++25}. The upper x-axis shows our measurement in terms of the projected mass in dark subhalos per unit area for halos with infall masses $10^6 < m/M_{\odot}< 10^{10.7}$. For a typical lens system with a deflector (source) redshift $z_{\rm{d}}=0.5$ ($z_{\rm{s}}=2.0$), we compute the projected mass density in subhalos, $\Sigma$, as (see Section VI C in G25)
\begin{equation}
\Sigma = 6.2 \times 10^6 \left(\frac{\Sigma_{\rm{sub}}}{0.1 \ \rm{kpc^{-2}}}\right) M_{\odot} \ \rm{kpc^{-2}}.
\end{equation}
Similarly, we relate $\Sigma_{\rm{sub}}$ to the projected mass fraction in subhalos near lensed images, $f_{\rm{sub}}$, according to 
\begin{equation}
f_{\rm{sub}} = 0.012 \left(\frac{\Sigma_{\rm{sub}}}{0.1 \ \rm{kpc^{-2}}}\right).
\end{equation}
At $95 \%$ confidence, we infer $\Sigma = 1.7_{-1.2}^{+2.6} \times 10^7 \ \mathrm{M}_{\odot} \rm{kpc^{-2}}$, or $f_{\rm{sub}} = 3.2_{-2.3}^{+5.1} \%$. This measurement is consistent at the $2 \sigma$ level with predictions from both {\tt{galacticus}} ($\Sigma = 0.9 \times 10^7 M_{\odot} \ \rm{kpc^{-2}}$) and Symphony ($\Sigma = 0.6 \times 10^7 M_{\odot} \ \rm{kpc^{-2}}$), marked as vertical green and magenta bars in Figure~\ref{fig:sigmasubinference}.

\textit{Discussion}---These results place among the strongest bounds on the free-streaming length of dark matter from a single independent inference, and yield the most precise measurement of subhalo abundance around strong lenses to date. We note a tentative $\sim 1.8 \sigma$ tension between our measurement of subhalo abundance and predictions from the Symphony $N$-body simulations \citep{Nadler++23,Gannon++25}. Additional data will reduce the statistical uncertainties of our measurement, and reveal if this tension is real. If it persists, it could indicate either that $N$-body simulations slightly under-predict subhalo abundance around strong lens systems, or that lenses include more sources of small-scale perturbation than expected. Possible factors that would lead us to infer higher subhalo abundance include an under estimation of the number of line of sight halos, or an under estimation of halo concentration. 

Our analysis provides a glimpse of the groundbreaking dark matter science that will become the norm in the forthcoming era of large cosmological surveys, during which time we will discover hundreds of new quadruply imaged quasars \citep{OguriMarshall10,Shajib++25}. We expect further improvements in our constraints on WDM models if dark matter is cold, or a detection of a turnover in the halo mass function, if dark matter is warm \citep{Gilman++24}. Looking beyond WDM, the techniques we present can be applied to any model that alters the abundance and internal structure of halos, relative to CDM. Notable examples include self-interacting dark matter, in which dark matter halos can undergo core collapse \citep{Gilman++21,Gilman++23,Enzi++25,Powell++25}, and modifications to the primordial matter power spectrum, which can increase both halo abundance and concentration \citep{ZentnerBullock03,Gilman++22,Nadler++25}. Finally, we note that the combination of strong lensing with other probes of small-scale structure, such as dwarf galaxies \citep{Nadler++21} and stellar streams \citep{Banik++21,Nibauer++25}, can break covariances present in each method when considered individually, leading to more precise inferences of halo abundance and internal structure \citep{Nadler+21b}.  

\section*{Acknowledgments}
    We thank Alex Drlica-Wagner, Josh Frieman, and Ethan Nadler for helpful discussions. We also thank the anonymous referees for constructive feedback that improved the quality of this work.
	
    DG acknowledges support for this work provided by the Brinson Foundation through a Brinson Prize Fellowship grant. 

    AMN, CG, MO and RK acknowledge support support from the National Science Foundation through the grant ``CAREER: An order of magnitude improvement in measurements of the physical properties of dark matter" NSF-AST-2442975.
    
    AMN, TT, XD, HP, CG, MO, RK acknowledge support from the National Science Foundation through the grant ``Collaborative Research: Measuring the physical properties of dark matter with strong gravitational lensing" NSF-AST-2205100, NSF-AST-2206315. 

    DW acknowledges support by NSF through grants NSF-AST-1906976 and NSF-AST-1836016, and from the Moore Foundation through grant 8548.

    D. Sluse acknowledges the support of the Fonds de la Recherche Scientifique-FNRS, Belgium, under grant No. 4.4503.1 and the Belgian Federal Science Policy Office (BELSPO) for the provision of financial support in the framework of the PRODEX Programme of the European Space Agency (ESA) under contract number 4000142531.

    P.M. acknowledges support from the National Science Foundation through grant NSF-AST-2407277. 

    SB acknowledges support by the Department of Physics and Astronomy, Stony Brook University

    TA acknowledges support from ANID-FONDECYT Regular Project 1240105 and the ANID BASAL project FB210003.

    KNA acknowledges support by the U.S. National Science Foundation (NSF) Theoretical Physics Program Grant No.\ PHY-2210283. 

    V.N.B. acknowledges funding support from STScI grant Nos. HST-GO-17103 and HST-AR-17063. 
    
    SGD acknowledges a generous support from the Ajax Foundation.

    SFH acknowledges support through UK Research and Innovation (UKRI) under the UK government’s Horizon Europe Funding Guarantee (EP/Z533920/1, selected in the 2023 ERC Advanced Grant round) and an STFC Small Award (ST/Y001656/1).
    
    A. K. was supported by the U.S. Department of Energy (DOE) Grant No. DE-SC0009937;  by World Premier International Research Center Initiative (WPI), MEXT, Japan; and by Japan Society for the Promotion of Science (JSPS) KAKENHI Grant No. JP20H05853.
    
    V.M. acknowledges support from ANID FONDECYT Regular grant number 1231418 and Centro de Astrof\'{\i}sica de Valpara\'{\i}so CIDI 21.

    Part of this work was carried out at the Jet Propulsion Laboratory, California Institute of Technology, under a contract with NASA.

    MS acknowledges partial support from NASA grant 80NSSC22K1294. 

    K.C.W. is supported by JSPS KAKENHI Grant Numbers JP24K07089, JP24H00221.
	
	This work is based on observations made with the James Webb Space Telescope through the Cycle 1 program JWST GO-2046 (PI:Nierenberg), and the Hubble Space Telescope through HST-GO-15320, HST-GO-15652, HST-GO-17916 (PI:Treu) and HST-GO-13732 (PI:Nierenberg). Funding from NASA through this programs is gratefully acknowledged.

    Some of the data presented herein were obtained at Keck Observatory, which is a private 501(c)3 non-profit organization operated as a scientific partnership among the California Institute of Technology, the University of California, and the National Aeronautics and Space Administration. The Keck facilities we used were LRIS and OSIRIS. The Observatory was made possible by the generous financial support of the W. M. Keck Foundation.  The authors wish to recognize and acknowledge the very significant cultural role and reverence that the summit of Maunakea has always had within the Native Hawaiian community. We are most fortunate to have the opportunity to conduct observations from this mountain.

    This research is based in part on data collected at the Subaru Telescope, which is operated by the National Astronomical Observatory of Japan. We are honored and grateful for the opportunity of observing the Universe from Maunakea, which has cultural, historical, and natural significance in Hawaii.

	This work used computational and storage services provided by the University of Chicago’s Research Computing Center; Caltech's Resnick High Performance Computing Center through Carnegie Science's partnership; the Pinnacles (NSF MRI, $\#$ 2019144) and CENVAL-ARC (NSF $\#$ 2346744) computing clusters at the Cyberinfrastructure and Research Technologies (CIRT) at University of California, Merced; and the Hoffman2 Cluster which is operated by the UCLA Office of Advanced Research Computing’s Research Technology Group.  
	
	\section*{Software} This work made use of {\tt{astropy}}:\footnote{\url{http://www.astropy.org}} a community-developed core Python package and an ecosystem of tools and resources for astronomy \citep{astropy:2013, astropy:2018, astropy:2022};  {\tt{cobyqa}} \citep{rago_thesis,razh_cobyqa}; {\tt{colossus}} \citep{Diemer18};  {\tt{lenstronomy}}\footnote{\url{https://github.com/lenstronomy/lenstronomy}} \citep{Birrer++18,Birrer++21}; {\tt{numpy}} \citep{numpy}; {\tt{pyHalo}}\footnote{\url{https://github.com/dangilman/pyHalo}} \citep{Gilman++20}; {\tt{trikde}}\footnote{\url{https://github.com/dangilman/trikde}}; {\tt{samana}}\footnote{\url{https://github.com/dangilman/samana}}; and {\tt{scipy}} \citep{scipy}. 

    \section*{Data availability}
	The data used in this article come from HST-GO-15320, HST-GO-15652, HST-GO-17917, HST-GO-13732 and JWST GO-2046 and GTO-1198. The raw data are publicly available online (\url{https://archive.stsci.edu/publishing/doi}). Astrometry and and flux ratio measurements are presented by \citet{Nierenberg++14,Nierenberg++20,Keeley++24,Keeley++25}. Reduced imaging data for the systems analyzed in this work are available in the open-source software {\tt{samana}}, which also provides notebooks that perform the lens modeling and scripts to reproduce the dark matter analysis.

\bibliography{apssamp}% Produces the bibliography via BibTeX.

@ARTICLE{Gavazzi_galaxy_density,
       author = {{Gavazzi}, Rapha{\"e}l and {Treu}, Tommaso and {Rhodes}, Jason D. and {Koopmans}, L{\'e}on V.~E. and {Bolton}, Adam S. and {Burles}, Scott and {Massey}, Richard J. and {Moustakas}, Leonidas A.},
        title = "{The Sloan Lens ACS Survey. IV. The Mass Density Profile of Early-Type Galaxies out to 100 Effective Radii}",
      journal = {\apj},
         year = 2007,
        month = sep,
       volume = {667},
       number = {1},
        pages = {176-190},
          doi = {10.1086/519237},
archivePrefix = {arXiv},
       eprint = {astro-ph/0701589},
 primaryClass = {astro-ph},
       adsurl = {https://ui.adsabs.harvard.edu/abs/2007ApJ...667..176G}
}

@article{Kusenko:2006rh,
    author = "Kusenko, Alexander",
    title = "{Sterile neutrinos, dark matter, and the pulsar velocities in models with a Higgs singlet}",
    eprint = "hep-ph/0609081",
    archivePrefix = "arXiv",
    reportNumber = "UCLA-06-TEP-23",
    doi = "10.1103/PhysRevLett.97.241301",
    journal = "Phys. Rev. Lett.",
    volume = "97",
    pages = "241301",
    year = "2006"
}

@article{Petraki:2007gq,
    author = "Petraki, Kalliopi and Kusenko, Alexander",
    title = "{Dark-matter sterile neutrinos in models with a gauge singlet in the Higgs sector}",
    eprint = "0711.4646",
    archivePrefix = "arXiv",
    primaryClass = "hep-ph",
    reportNumber = "UCLA-07-TEP-27",
    doi = "10.1103/PhysRevD.77.065014",
    journal = "Phys. Rev. D",
    volume = "77",
    pages = "065014",
    year = "2008"
}

@article{Petraki:2008ef,
    author = "Petraki, Kalliopi",
    title = "{Small-scale structure formation properties of chilled sterile neutrinos as dark matter}",
    eprint = "0801.3470",
    archivePrefix = "arXiv",
    primaryClass = "hep-ph",
    reportNumber = "UCLA-08-TEP-02",
    doi = "10.1103/PhysRevD.77.105004",
    journal = "Phys. Rev. D",
    volume = "77",
    pages = "105004",
    year = "2008"
}

@article{Wambsganss_microlensing,
  title = {Gravitational Microlensing},
  author = {Wambsganss, Joachim},
  year = 2006,
  journal = {arXiv e-prints},
  number = {astro-ph/0604278},
  eprint = {astro-ph/0604278},
  pages = {astro-ph/0604278},
  doi = {10.48550/arXiv.astro-ph/0604278},
  archiveprefix = {arXiv}
}

@ARTICLE{Powell++25,
	author = {{Powell}, D.~M. and {McKean}, J.~P. and {Vegetti}, S. and {Spingola}, C. and {White}, S.~D.~M. and {Fassnacht}, C.~D.},
	title = "{A million-solar-mass object detected at a cosmological distance using gravitational imaging}",
	journal = {Nature Astronomy},
	year = 2025,
	month = oct,
	doi = {10.1038/s41550-025-02651-2},
	adsurl = {https://ui.adsabs.harvard.edu/abs/2025NatAs.tmp..205P}
}

@ARTICLE{Nibauer++25,
	author = {{Nibauer}, Jacob and {Bonaca}, Ana and {Price-Whelan}, Adrian M. and {Spergel}, David N. and {Greene}, Jenny E.},
	title = "{Measurement of Dark Matter Substructure from the Kinematics of the GD-1 Stellar Stream}",
	journal = {arXiv e-prints},
	year = 2025,
	month = oct,
	eid = {arXiv:2510.02247},
	pages = {arXiv:2510.02247},
	doi = {10.48550/arXiv.2510.02247},
	archivePrefix = {arXiv},
	eprint = {2510.02247},
	primaryClass = {astro-ph.GA},
	adsurl = {https://ui.adsabs.harvard.edu/abs/2025arXiv251002247N}
}

@ARTICLE{Navarro++97,
	author = {{Navarro}, Julio F. and {Frenk}, Carlos S. and {White}, Simon D.~M.},
	title = "{A Universal Density Profile from Hierarchical Clustering}",
	journal = {\apj},
	year = 1997,
	month = dec,
	volume = {490},
	number = {2},
	pages = {493-508},
	doi = {10.1086/304888},
	archivePrefix = {arXiv},
	eprint = {astro-ph/9611107},
	primaryClass = {astro-ph},
	adsurl = {https://ui.adsabs.harvard.edu/abs/1997ApJ...490..493N}
}

@ARTICLE{Nierenberg++17,
	author = {{Nierenberg}, A.~M. and {Treu}, T. and {Brammer}, G. and {Peter}, A.~H.~G. and {Fassnacht}, C.~D. and {Keeton}, C.~R. and {Kochanek}, C.~S. and {Schmidt}, K.~B. and {Sluse}, D. and {Wright}, S.~A.},
	title = "{Probing dark matter substructure in the gravitational lens HE 0435-1223 with the WFC3 grism}",
	journal = {\mnras},
	year = 2017,
	month = oct,
	volume = {471},
	number = {2},
	pages = {2224-2236},
	doi = {10.1093/mnras/stx1400},
	archivePrefix = {arXiv},
	eprint = {1701.05188},
	primaryClass = {astro-ph.CO},
	adsurl = {https://ui.adsabs.harvard.edu/abs/2017MNRAS.471.2224N}
}

@ARTICLE{Birrer++21,
	author = {{Birrer}, Simon and {Shajib}, Anowar and {Gilman}, Daniel and {Galan}, Aymeric and {Aalbers}, Jelle and {Millon}, Martin and {Morgan}, Robert and {Pagano}, Giulia and {Park}, Ji and {Teodori}, Luca and {Tessore}, Nicolas and {Ueland}, Madison and {Van de Vyvere}, Lyne and {Wagner-Carena}, Sebastian and {Wempe}, Ewoud and {Yang}, Lilan and {Ding}, Xuheng and {Schmidt}, Thomas and {Sluse}, Dominique and {Zhang}, Ming and {Amara}, Adam},
	title = "{lenstronomy II: A gravitational lensing software ecosystem}",
	journal = {The Journal of Open Source Software},
	year = 2021,
	month = jun,
	volume = {6},
	number = {62},
	eid = {3283},
	pages = {3283},
	doi = {10.21105/joss.03283},
	archivePrefix = {arXiv},
	eprint = {2106.05976},
	primaryClass = {astro-ph.CO},
	adsurl = {https://ui.adsabs.harvard.edu/abs/2021JOSS....6.3283B}
}

@ARTICLE{Birrer++18,
	author = {{Birrer}, Simon and {Amara}, Adam},
	title = "{lenstronomy: Multi-purpose gravitational lens modelling software package}",
	journal = {Physics of the Dark Universe},
	year = 2018,
	month = dec,
	volume = {22},
	pages = {189-201},
	doi = {10.1016/j.dark.2018.11.002},
	archivePrefix = {arXiv},
	eprint = {1803.09746},
	primaryClass = {astro-ph.CO},
	adsurl = {https://ui.adsabs.harvard.edu/abs/2018PDU....22..189B}
}

@ARTICLE{Enzi++25,
	author = {{Enzi}, Wolfgang J.~R. and {Krawczyk}, Coleman M. and {Ballard}, Daniel J. and {Collett}, Thomas E.},
	title = "{The overconcentrated dark halo in the strong lens SDSS J0946 + 1006 is a subhalo: evidence for self-interacting dark matter?}",
	journal = {\mnras},
	year = 2025,
	month = jun,
	volume = {540},
	number = {1},
	pages = {247-263},
	doi = {10.1093/mnras/staf697},
	archivePrefix = {arXiv},
	eprint = {2411.08565},
	primaryClass = {astro-ph.CO},
	adsurl = {https://ui.adsabs.harvard.edu/abs/2025MNRAS.540..247E}
}

@ARTICLE{Cohen++24,
	author = {{Cohen}, Jacob S. and {Fassnacht}, Christopher D. and {O'Riordan}, Conor M. and {Vegetti}, Simona},
	title = "{General multipoles and their implications for dark matter inference}",
	journal = {\mnras},
	year = 2024,
	month = jul,
	volume = {531},
	number = {3},
	pages = {3431-3443},
	doi = {10.1093/mnras/stae1228},
	archivePrefix = {arXiv},
	eprint = {2403.08895},
	primaryClass = {astro-ph.CO},
	adsurl = {https://ui.adsabs.harvard.edu/abs/2024MNRAS.531.3431C}
}

@ARTICLE{ZentnerBullock03,
	author = {{Zentner}, Andrew R. and {Bullock}, James S.},
	title = "{Halo Substructure and the Power Spectrum}",
	journal = {\apj},
	year = 2003,
	month = nov,
	volume = {598},
	number = {1},
	pages = {49-72},
	doi = {10.1086/378797},
	archivePrefix = {arXiv},
	eprint = {astro-ph/0304292},
	primaryClass = {astro-ph},
	adsurl = {https://ui.adsabs.harvard.edu/abs/2003ApJ...598...49Z}
}

@ARTICLE{AvilaReese++01,
	author = {{Avila-Reese}, Vladimir and {Col{\'\i}n}, Pedro and {Valenzuela}, Octavio and {D'Onghia}, Elena and {Firmani}, Claudio},
	title = "{Formation and Structure of Halos in a Warm Dark Matter Cosmology}",
	journal = {\apj},
	year = 2001,
	month = oct,
	volume = {559},
	number = {2},
	pages = {516-530},
	doi = {10.1086/322411},
	archivePrefix = {arXiv},
	eprint = {astro-ph/0010525},
	primaryClass = {astro-ph},
	adsurl = {https://ui.adsabs.harvard.edu/abs/2001ApJ...559..516A}
}

@article{leftley_parsec-scale_2019,
  title = {Parsec-Scale {{Dusty Winds}} in {{Active Galactic Nuclei}}: {{Evidence}} for {{Radiation Pressure Driving}}},
  shorttitle = {Parsec-Scale {{Dusty Winds}} in {{Active Galactic Nuclei}}},
  author = {Leftley, James H. and H{\"o}nig, Sebastian F. and Asmus, Daniel and Tristram, Konrad R. W. and Gandhi, Poshak and Kishimoto, Makoto and Venanzi, Marta and Williamson, David J.},
  year = 2019,
  journal = {ApJ},
  volume = {886},
  pages = {55},
  issn = {0004-637X},
  doi = {10.3847/1538-4357/ab4a0b}
}

@article{burtscher_diversity_2013,
  title = {A Diversity of Dusty {{AGN}} Tori. {{Data}} Release for the {{VLTI}}/{{MIDI AGN Large Program}} and First Results for 23 Galaxies},
  author = {Burtscher, L. and Meisenheimer, K. and Tristram, K. R. W. and Jaffe, W. and H{\"o}nig, S. F. and Davies, R. I. and Kishimoto, M. and Pott, J. -U. and R{\"o}ttgering, H. and Schartmann, M. and Weigelt, G. and Wolf, S.},
  year = 2013,
  journal = {Astronomy and Astrophysics},
  volume = {558},
  pages = {A149},
  issn = {0004-6361},
  doi = {10.1051/0004-6361/201321890}
}

@ARTICLE{Bond++83,
	author = {{Bond}, J.~R. and {Szalay}, A.~S.},
	title = "{The collisionless damping of density fluctuations in an expanding universe}",
	journal = {\apj},
	year = 1983,
	month = nov,
	volume = {274},
	pages = {443-468},
	doi = {10.1086/161460},
	adsurl = {https://ui.adsabs.harvard.edu/abs/1983ApJ...274..443B}
}

@ARTICLE{Schneider++13,
	author = {{Schneider}, Aurel and {Smith}, Robert E. and {Reed}, Darren},
	title = "{Halo mass function and the free streaming scale}",
	journal = {\mnras},
	year = 2013,
	month = aug,
	volume = {433},
	number = {2},
	pages = {1573-1587},
	doi = {10.1093/mnras/stt829},
	archivePrefix = {arXiv},
	eprint = {1303.0839},
	primaryClass = {astro-ph.CO},
	adsurl = {https://ui.adsabs.harvard.edu/abs/2013MNRAS.433.1573S}
}

@ARTICLE{BuckleyPeter18,
       author = {{Buckley}, Matthew R. and {Peter}, Annika H.~G.},
        title = "{Gravitational probes of dark matter physics}",
      journal = {\physrep},
         year = 2018,
        month = oct,
       volume = {761},
        pages = {1-60},
          doi = {10.1016/j.physrep.2018.07.003},
archivePrefix = {arXiv},
       eprint = {1712.06615},
 primaryClass = {astro-ph.CO},
       adsurl = {https://ui.adsabs.harvard.edu/abs/2018PhR...761....1B}
}

@ARTICLE{Nadler+21b,
       author = {{Nadler}, Ethan O. and {Birrer}, Simon and {Gilman}, Daniel and {Wechsler}, Risa H. and {Du}, Xiaolong and {Benson}, Andrew and {Nierenberg}, Anna M. and {Treu}, Tommaso},
        title = "{Dark Matter Constraints from a Unified Analysis of Strong Gravitational Lenses and Milky Way Satellite Galaxies}",
      journal = {\apj},
         year = 2021,
        month = aug,
       volume = {917},
       number = {1},
          eid = {7},
        pages = {7},
          doi = {10.3847/1538-4357/abf9a3},
archivePrefix = {arXiv},
       eprint = {2101.07810},
 primaryClass = {astro-ph.CO},
       adsurl = {https://ui.adsabs.harvard.edu/abs/2021ApJ...917....7N}
}

@ARTICLE{Despali++18,
       author = {{Despali}, Giulia and {Vegetti}, Simona and {White}, Simon D.~M. and {Giocoli}, Carlo and {van den Bosch}, Frank C.},
        title = "{Modelling the line-of-sight contribution in substructure lensing}",
      journal = {\mnras},
         year = 2018,
        month = apr,
       volume = {475},
       number = {4},
        pages = {5424-5442},
          doi = {10.1093/mnras/sty159},
archivePrefix = {arXiv},
       eprint = {1710.05029},
 primaryClass = {astro-ph.CO},
       adsurl = {https://ui.adsabs.harvard.edu/abs/2018MNRAS.475.5424D}
}

@ARTICLE{Ludlow++16,
       author = {{Ludlow}, Aaron D. and {Bose}, Sownak and {Angulo}, Ra{\'u}l E. and {Wang}, Lan and {Hellwing}, Wojciech A. and {Navarro}, Julio F. and {Cole}, Shaun and {Frenk}, Carlos S.},
        title = "{The mass-concentration-redshift relation of cold and warm dark matter haloes}",
      journal = {\mnras},
         year = 2016,
        month = aug,
       volume = {460},
       number = {2},
        pages = {1214-1232},
          doi = {10.1093/mnras/stw1046},
archivePrefix = {arXiv},
       eprint = {1601.02624},
 primaryClass = {astro-ph.CO},
       adsurl = {https://ui.adsabs.harvard.edu/abs/2016MNRAS.460.1214L}
}

@ARTICLE{Boddy++22,
       author = {{Boddy}, Kimberly K. and {Lisanti}, Mariangela and {McDermott}, Samuel D. and {Rodd}, Nicholas L. and {Weniger}, Christoph and {Ali-Ha{\"\i}moud}, Yacine and {Buschmann}, Malte and {Cholis}, Ilias and {Croon}, Djuna and {Erickcek}, Adrienne L. and {Gluscevic}, Vera and {Leane}, Rebecca K. and {Mishra-Sharma}, Siddharth and {Mu{\~n}oz}, Julian B. and {Nadler}, Ethan O. and {Natarajan}, Priyamvada and {Price-Whelan}, Adrian and {Vegetti}, Simona and {Witte}, Samuel J.},
        title = "{Snowmass2021 theory frontier white paper: Astrophysical and cosmological probes of dark matter}",
      journal = {Journal of High Energy Astrophysics},
         year = 2022,
        month = aug,
       volume = {35},
        pages = {112-138},
          doi = {10.1016/j.jheap.2022.06.005},
archivePrefix = {arXiv},
       eprint = {2203.06380},
 primaryClass = {hep-ph},
       adsurl = {https://ui.adsabs.harvard.edu/abs/2022JHEAp..35..112B}
}

@ARTICLE{DrlicaWagner22,
	author = {{Drlica-Wagner}, Alex and {Prescod-Weinstein}, Chanda and {Yu}, Hai-Bo and {Albert}, Andrea and {Amin}, Mustafa and {Banerjee}, Arka and {Baryakhtar}, Masha and {Bechtol}, Keith and {Bird}, Simeon and {Birrer}, Simon and {Bringmann}, Torsten and {Caputo}, Regina and {Chakrabarti}, Sukanya and {Chen}, Thomas Y. and {Croon}, Djuna and {Cyr-Racine}, Francis-Yan and {Dawson}, William A. and {Dvorkin}, Cora and {Gluscevic}, Vera and {Gilman}, Daniel and {Grin}, Daniel and {Hlo{\v{z}}ek}, Ren{\'e}e and {Leane}, Rebecca K. and {Li}, Ting S. and {Mao}, Yao-Yuan and {Meyers}, Joel and {Mishra-Sharma}, Siddharth and {Mu{\~n}oz}, Julian B. and {Munshi}, Ferah and {Nadler}, Ethan O. and {Parikh}, Aditya and {Perez}, Kerstin and {Peter}, Annika H.~G. and {Profumo}, Stefano and {Schutz}, Katelin and {Sehgal}, Neelima and {Simon}, Joshua D. and {Sinha}, Kuver and {Valluri}, Monica and {Wechsler}, Risa H.},
	title = "{Report of the Topical Group on Cosmic Probes of Dark Matter for Snowmass 2021}",
	journal = {arXiv e-prints},
	year = 2022,
	month = sep,
	eid = {arXiv:2209.08215},
	pages = {arXiv:2209.08215},
	doi = {10.48550/arXiv.2209.08215},
	archivePrefix = {arXiv},
	eprint = {2209.08215},
	primaryClass = {hep-ph},
	adsurl = {https://ui.adsabs.harvard.edu/abs/2022arXiv220908215D}
}

@ARTICLE{MoustakasMetcalf03,
	author = {{Moustakas}, Leonidas A. and {Metcalf}, R. Benton},
	title = "{Detecting dark matter substructure spectroscopically in strong gravitational lenses}",
	journal = {\mnras},
	year = 2003,
	month = mar,
	volume = {339},
	number = {3},
	pages = {607-615},
	doi = {10.1046/j.1365-8711.2003.06055.x},
	archivePrefix = {arXiv},
	eprint = {astro-ph/0206176},
	primaryClass = {astro-ph},
	adsurl = {https://ui.adsabs.harvard.edu/abs/2003MNRAS.339..607M}
}

@ARTICLE{Auger++10,
	author = {{Auger}, M.~W. and {Treu}, T. and {Bolton}, A.~S. and {Gavazzi}, R. and {Koopmans}, L.~V.~E. and {Marshall}, P.~J. and {Moustakas}, L.~A. and {Burles}, S.},
	title = "{The Sloan Lens ACS Survey. X. Stellar, Dynamical, and Total Mass Correlations of Massive Early-type Galaxies}",
	journal = {\apj},
	year = 2010,
	month = nov,
	volume = {724},
	number = {1},
	pages = {511-525},
	doi = {10.1088/0004-637X/724/1/511},
	archivePrefix = {arXiv},
	eprint = {1007.2880},
	primaryClass = {astro-ph.CO},
	adsurl = {https://ui.adsabs.harvard.edu/abs/2010ApJ...724..511A}
}

@ARTICLE{Dalal++02,
	author = {{Dalal}, N. and {Kochanek}, C.~S.},
	title = "{Direct Detection of Cold Dark Matter Substructure}",
	journal = {\apj},
	year = 2002,
	month = jun,
	volume = {572},
	number = {1},
	pages = {25-33},
	doi = {10.1086/340303},
	archivePrefix = {arXiv},
	eprint = {astro-ph/0111456},
	primaryClass = {astro-ph},
	adsurl = {https://ui.adsabs.harvard.edu/abs/2002ApJ...572...25D}
}

@ARTICLE{Hezaveh++16,
	author = {{Hezaveh}, Yashar D. and {Dalal}, Neal and {Marrone}, Daniel P. and {Mao}, Yao-Yuan and {Morningstar}, Warren and {Wen}, Di and {Blandford}, Roger D. and {Carlstrom}, John E. and {Fassnacht}, Christopher D. and {Holder}, Gilbert P. and {Kemball}, Athol and {Marshall}, Philip J. and {Murray}, Norman and {Perreault Levasseur}, Laurence and {Vieira}, Joaquin D. and {Wechsler}, Risa H.},
	title = "{Detection of Lensing Substructure Using ALMA Observations of the Dusty Galaxy SDP.81}",
	journal = {\apj},
	year = 2016,
	month = may,
	volume = {823},
	number = {1},
	eid = {37},
	pages = {37},
	doi = {10.3847/0004-637X/823/1/37},
	archivePrefix = {arXiv},
	eprint = {1601.01388},
	primaryClass = {astro-ph.CO},
	adsurl = {https://ui.adsabs.harvard.edu/abs/2016ApJ...823...37H}
}

@ARTICLE{Viel++05,
       author = {{Viel}, Matteo and {Lesgourgues}, Julien and {Haehnelt}, Martin G. and {Matarrese}, Sabino and {Riotto}, Antonio},
        title = "{Constraining warm dark matter candidates including sterile neutrinos and light gravitinos with WMAP and the Lyman-{\ensuremath{\alpha}} forest}",
      journal = {\prd},
         year = 2005,
        month = mar,
       volume = {71},
       number = {6},
          eid = {063534},
        pages = {063534},
          doi = {10.1103/PhysRevD.71.063534},
archivePrefix = {arXiv},
       eprint = {astro-ph/0501562},
 primaryClass = {astro-ph},
       adsurl = {https://ui.adsabs.harvard.edu/abs/2005PhRvD..71f3534V}
}

@ARTICLE{Planck++20,
	author = {{Planck Collaboration} and {Aghanim}, N. and {Akrami}, Y. and {Ashdown}, M. and {Aumont}, J. and {Baccigalupi}, C. and {Ballardini}, M. and {Banday}, A.~J. and {Barreiro}, R.~B. and {Bartolo}, N. and {Basak}, S. and {Battye}, R. and {Benabed}, K. and {Bernard}, J. -P. and {Bersanelli}, M. and {Bielewicz}, P. and {Bock}, J.~J. and {Bond}, J.~R. and {Borrill}, J. and {Bouchet}, F.~R. and {Boulanger}, F. and {Bucher}, M. and {Burigana}, C. and {Butler}, R.~C. and {Calabrese}, E. and {Cardoso}, J. -F. and {Carron}, J. and {Challinor}, A. and {Chiang}, H.~C. and {Chluba}, J. and {Colombo}, L.~P.~L. and {Combet}, C. and {Contreras}, D. and {Crill}, B.~P. and {Cuttaia}, F. and {de Bernardis}, P. and {de Zotti}, G. and {Delabrouille}, J. and {Delouis}, J. -M. and {Di Valentino}, E. and {Diego}, J.~M. and {Dor{\'e}}, O. and {Douspis}, M. and {Ducout}, A. and {Dupac}, X. and {Dusini}, S. and {Efstathiou}, G. and {Elsner}, F. and {En{\ss}lin}, T.~A. and {Eriksen}, H.~K. and {Fantaye}, Y. and {Farhang}, M. and {Fergusson}, J. and {Fernandez-Cobos}, R. and {Finelli}, F. and {Forastieri}, F. and {Frailis}, M. and {Fraisse}, A.~A. and {Franceschi}, E. and {Frolov}, A. and {Galeotta}, S. and {Galli}, S. and {Ganga}, K. and {G{\'e}nova-Santos}, R.~T. and {Gerbino}, M. and {Ghosh}, T. and {Gonz{\'a}lez-Nuevo}, J. and {G{\'o}rski}, K.~M. and {Gratton}, S. and {Gruppuso}, A. and {Gudmundsson}, J.~E. and {Hamann}, J. and {Handley}, W. and {Hansen}, F.~K. and {Herranz}, D. and {Hildebrandt}, S.~R. and {Hivon}, E. and {Huang}, Z. and {Jaffe}, A.~H. and {Jones}, W.~C. and {Karakci}, A. and {Keih{\"a}nen}, E. and {Keskitalo}, R. and {Kiiveri}, K. and {Kim}, J. and {Kisner}, T.~S. and {Knox}, L. and {Krachmalnicoff}, N. and {Kunz}, M. and {Kurki-Suonio}, H. and {Lagache}, G. and {Lamarre}, J. -M. and {Lasenby}, A. and {Lattanzi}, M. and {Lawrence}, C.~R. and {Le Jeune}, M. and {Lemos}, P. and {Lesgourgues}, J. and {Levrier}, F. and {Lewis}, A. and {Liguori}, M. and {Lilje}, P.~B. and {Lilley}, M. and {Lindholm}, V. and {L{\'o}pez-Caniego}, M. and {Lubin}, P.~M. and {Ma}, Y. -Z. and {Mac{\'\i}as-P{\'e}rez}, J.~F. and {Maggio}, G. and {Maino}, D. and {Mandolesi}, N. and {Mangilli}, A. and {Marcos-Caballero}, A. and {Maris}, M. and {Martin}, P.~G. and {Martinelli}, M. and {Mart{\'\i}nez-Gonz{\'a}lez}, E. and {Matarrese}, S. and {Mauri}, N. and {McEwen}, J.~D. and {Meinhold}, P.~R. and {Melchiorri}, A. and {Mennella}, A. and {Migliaccio}, M. and {Millea}, M. and {Mitra}, S. and {Miville-Desch{\^e}nes}, M. -A. and {Molinari}, D. and {Montier}, L. and {Morgante}, G. and {Moss}, A. and {Natoli}, P. and {N{\o}rgaard-Nielsen}, H.~U. and {Pagano}, L. and {Paoletti}, D. and {Partridge}, B. and {Patanchon}, G. and {Peiris}, H.~V. and {Perrotta}, F. and {Pettorino}, V. and {Piacentini}, F. and {Polastri}, L. and {Polenta}, G. and {Puget}, J. -L. and {Rachen}, J.~P. and {Reinecke}, M. and {Remazeilles}, M. and {Renzi}, A. and {Rocha}, G. and {Rosset}, C. and {Roudier}, G. and {Rubi{\~n}o-Mart{\'\i}n}, J.~A. and {Ruiz-Granados}, B. and {Salvati}, L. and {Sandri}, M. and {Savelainen}, M. and {Scott}, D. and {Shellard}, E.~P.~S. and {Sirignano}, C. and {Sirri}, G. and {Spencer}, L.~D. and {Sunyaev}, R. and {Suur-Uski}, A. -S. and {Tauber}, J.~A. and {Tavagnacco}, D. and {Tenti}, M. and {Toffolatti}, L. and {Tomasi}, M. and {Trombetti}, T. and {Valenziano}, L. and {Valiviita}, J. and {Van Tent}, B. and {Vibert}, L. and {Vielva}, P. and {Villa}, F. and {Vittorio}, N. and {Wandelt}, B.~D. and {Wehus}, I.~K. and {White}, M. and {White}, S.~D.~M. and {Zacchei}, A. and {Zonca}, A.},
	title = "{Planck 2018 results. VI. Cosmological parameters}",
	journal = {\aap},
	year = 2020,
	month = sep,
	volume = {641},
	eid = {A6},
	pages = {A6},
	doi = {10.1051/0004-6361/201833910},
	archivePrefix = {arXiv},
	eprint = {1807.06209},
	primaryClass = {astro-ph.CO},
	adsurl = {https://ui.adsabs.harvard.edu/abs/2020A&A...641A...6P}
}

@ARTICLE{OguriMarshall10,
	author = {{Oguri}, Masamune and {Marshall}, Philip J.},
	title = "{Gravitationally lensed quasars and supernovae in future wide-field optical imaging surveys}",
	journal = {\mnras},
	year = 2010,
	month = jul,
	volume = {405},
	number = {4},
	pages = {2579-2593},
	doi = {10.1111/j.1365-2966.2010.16639.x},
	archivePrefix = {arXiv},
	eprint = {1001.2037},
	primaryClass = {astro-ph.CO},
	adsurl = {https://ui.adsabs.harvard.edu/abs/2010MNRAS.405.2579O}
}

@ARTICLE{Shajib++25,
	author = {{Shajib}, Anowar J. and {Smith}, Graham P. and {Birrer}, Simon and {Verma}, Aprajita and {Arendse}, Nikki and {Collett}, Thomas and {Daylan}, Tansu and {Serjeant}, Stephen and {LSST Strong Lensing Science Collaboration}},
	title = "{Strong gravitational lenses from the Vera C. Rubin Observatory}",
	journal = {Philosophical Transactions of the Royal Society of London Series A},
	year = 2025,
	month = may,
	volume = {383},
	number = {2295},
	eid = {20240117},
	pages = {20240117},
	doi = {10.1098/rsta.2024.0117},
	archivePrefix = {arXiv},
	eprint = {2406.08919},
	primaryClass = {astro-ph.CO},
	adsurl = {https://ui.adsabs.harvard.edu/abs/2025RSPTA.38340117S}
}

@ARTICLE{Nadler++25,
	author = {{Nadler}, Ethan O. and {Gluscevic}, Vera and {Benson}, Andrew},
	title = "{The Effects of Linear Matter Power Spectrum Enhancement on Dark Matter Substructure}",
	journal = {arXiv e-prints},
	year = 2025,
	month = jul,
	eid = {arXiv:2507.16889},
	pages = {arXiv:2507.16889},
	doi = {10.48550/arXiv.2507.16889},
	archivePrefix = {arXiv},
	eprint = {2507.16889},
	primaryClass = {astro-ph.CO},
	adsurl = {https://ui.adsabs.harvard.edu/abs/2025arXiv250716889N}
}

@ARTICLE{MaoSchneider98,
	author = {{Mao}, Shude and {Schneider}, Peter},
	title = "{Evidence for substructure in lens galaxies?}",
	journal = {\mnras},
	year = 1998,
	month = apr,
	volume = {295},
	number = {3},
	pages = {587-594},
	doi = {10.1046/j.1365-8711.1998.01319.x},
	archivePrefix = {arXiv},
	eprint = {astro-ph/9707187},
	primaryClass = {astro-ph},
	adsurl = {https://ui.adsabs.harvard.edu/abs/1998MNRAS.295..587M}
}

@ARTICLE{Banik++21,
	author = {{Banik}, Nilanjan and {Bovy}, Jo and {Bertone}, Gianfranco and {Erkal}, Denis and {de Boer}, T.~J.~L.},
	title = "{Novel constraints on the particle nature of dark matter from stellar streams}",
	journal = {\jcap},
	year = 2021,
	month = oct,
	volume = {2021},
	number = {10},
	eid = {043},
	pages = {043},
	doi = {10.1088/1475-7516/2021/10/043},
	archivePrefix = {arXiv},
	eprint = {1911.02663},
	primaryClass = {astro-ph.GA},
	adsurl = {https://ui.adsabs.harvard.edu/abs/2021JCAP...10..043B}
}

@ARTICLE{Nadler++21,
	author = {{Nadler}, E.~O. and {Drlica-Wagner}, A. and {Bechtol}, K. and {Mau}, S. and {Wechsler}, R.~H. and {Gluscevic}, V. and {Boddy}, K. and {Pace}, A.~B. and {Li}, T.~S. and {McNanna}, M. and {Riley}, A.~H. and {Garc{\'\i}a-Bellido}, J. and {Mao}, Y. -Y. and {Green}, G. and {Burke}, D.~L. and {Peter}, A. and {Jain}, B. and {Abbott}, T.~M.~C. and {Aguena}, M. and {Allam}, S. and {Annis}, J. and {Avila}, S. and {Brooks}, D. and {Carrasco Kind}, M. and {Carretero}, J. and {Costanzi}, M. and {da Costa}, L.~N. and {De Vicente}, J. and {Desai}, S. and {Diehl}, H.~T. and {Doel}, P. and {Everett}, S. and {Evrard}, A.~E. and {Flaugher}, B. and {Frieman}, J. and {Gerdes}, D.~W. and {Gruen}, D. and {Gruendl}, R.~A. and {Gschwend}, J. and {Gutierrez}, G. and {Hinton}, S.~R. and {Honscheid}, K. and {Huterer}, D. and {James}, D.~J. and {Krause}, E. and {Kuehn}, K. and {Kuropatkin}, N. and {Lahav}, O. and {Maia}, M.~A.~G. and {Marshall}, J.~L. and {Menanteau}, F. and {Miquel}, R. and {Palmese}, A. and {Paz-Chinch{\'o}n}, F. and {Plazas}, A.~A. and {Romer}, A.~K. and {Sanchez}, E. and {Scarpine}, V. and {Serrano}, S. and {Sevilla-Noarbe}, I. and {Smith}, M. and {Soares-Santos}, M. and {Suchyta}, E. and {Swanson}, M.~E.~C. and {Tarle}, G. and {Tucker}, D.~L. and {Walker}, A.~R. and {Wester}, W. and {DES Collaboration}},
	title = "{Constraints on Dark Matter Properties from Observations of Milky Way Satellite Galaxies}",
	journal = {\prl},
	year = 2021,
	month = mar,
	volume = {126},
	number = {9},
	eid = {091101},
	pages = {091101},
	doi = {10.1103/PhysRevLett.126.091101},
	archivePrefix = {arXiv},
	eprint = {2008.00022},
	primaryClass = {astro-ph.CO},
	adsurl = {https://ui.adsabs.harvard.edu/abs/2021PhRvL.126i1101N}
}

@ARTICLE{Zelko++22,
       author = {{Zelko}, Ioana A. and {Treu}, Tommaso and {Abazajian}, Kevork N. and {Gilman}, Daniel and {Benson}, Andrew J. and {Birrer}, Simon and {Nierenberg}, Anna M. and {Kusenko}, Alexander},
        title = "{Constraints on Sterile Neutrino Models from Strong Gravitational Lensing, Milky Way Satellites, and the Lyman-{\ensuremath{\alpha}} Forest}",
      journal = {\prl},
         year = 2022,
        month = nov,
       volume = {129},
       number = {19},
          eid = {191301},
        pages = {191301},
          doi = {10.1103/PhysRevLett.129.191301},
archivePrefix = {arXiv},
       eprint = {2205.09777},
 primaryClass = {hep-ph},
       adsurl = {https://ui.adsabs.harvard.edu/abs/2022PhRvL.129s1301Z}
}

@ARTICLE{Birrer++17,
	author = {{Birrer}, Simon and {Amara}, Adam and {Refregier}, Alexandre},
	title = "{Lensing substructure quantification in RXJ1131-1231: a 2 keV lower bound on dark matter thermal relic mass}",
	journal = {\jcap},
	year = 2017,
	month = may,
	volume = {2017},
	number = {5},
	eid = {037},
	pages = {037},
	doi = {10.1088/1475-7516/2017/05/037},
	archivePrefix = {arXiv},
	eprint = {1702.00009},
	primaryClass = {astro-ph.CO},
	adsurl = {https://ui.adsabs.harvard.edu/abs/2017JCAP...05..037B}
}

@article{honig_redefining_2019,
  title = {Redefining the {{Torus}}: {{A Unifying View}} of {{AGNs}} in the {{Infrared}} and {{Submillimeter}}},
  shorttitle = {Redefining the {{Torus}}},
  author = {H{\"o}nig, Sebastian F.},
  year = 2019,
  journal = {ApJ},
  volume = {884},
  pages = {171},
  issn = {0004-637X},
  doi = {10.3847/1538-4357/ab4591}
}

@ARTICLE{Muller-Sanchez++11,
	author = {{M{\"u}ller-S{\'a}nchez}, F. and {Prieto}, M.~A. and {Hicks}, E.~K.~S. and {Vives-Arias}, H. and {Davies}, R.~I. and {Malkan}, M. and {Tacconi}, L.~J. and {Genzel}, R.},
	title = "{Outflows from Active Galactic Nuclei: Kinematics of the Narrow-line and Coronal-line Regions in Seyfert Galaxies}",
	journal = {\apj},
	year = 2011,
	month = oct,
	volume = {739},
	number = {2},
	eid = {69},
	pages = {69},
	doi = {10.1088/0004-637X/739/2/69},
	archivePrefix = {arXiv},
	eprint = {1107.3140},
	primaryClass = {astro-ph.CO},
	adsurl = {https://ui.adsabs.harvard.edu/abs/2011ApJ...739...69M}
}

@ARTICLE{Dobler++06,
	author = {{Dobler}, Gregory and {Keeton}, Charles R.},
	title = "{Finite source effects in strong lensing: implications for the substructure mass scale}",
	journal = {\mnras},
	year = 2006,
	month = feb,
	volume = {365},
	number = {4},
	pages = {1243-1262},
	doi = {10.1111/j.1365-2966.2005.09809.x},
	archivePrefix = {arXiv},
	eprint = {astro-ph/0502436},
	primaryClass = {astro-ph},
	adsurl = {https://ui.adsabs.harvard.edu/abs/2006MNRAS.365.1243D}
}

@ARTICLE{Oh++24,
	author = {{Oh}, Maverick S.~H. and {Nierenberg}, Anna and {Gilman}, Daniel and {Birrer}, Simon},
	title = "{Joint Semi-Analytic Multipole Priors from Galaxy Isophotes and Constraints from Lensed Arcs}",
	journal = {arXiv e-prints},
	year = 2024,
	month = apr,
	eid = {arXiv:2404.17124},
	pages = {arXiv:2404.17124},
	doi = {10.48550/arXiv.2404.17124},
	archivePrefix = {arXiv},
	eprint = {2404.17124},
	primaryClass = {astro-ph.CO},
	adsurl = {https://ui.adsabs.harvard.edu/abs/2024arXiv240417124O}
}

@ARTICLE{Nadler++23,
	author = {{Nadler}, Ethan O. and {Mansfield}, Philip and {Wang}, Yunchong and {Du}, Xiaolong and {Adhikari}, Susmita and {Banerjee}, Arka and {Benson}, Andrew and {Darragh-Ford}, Elise and {Mao}, Yao-Yuan and {Wagner-Carena}, Sebastian and {Wechsler}, Risa H. and {Wu}, Hao-Yi},
	title = "{Symphony: Cosmological Zoom-in Simulation Suites over Four Decades of Host Halo Mass}",
	journal = {\apj},
	year = 2023,
	month = mar,
	volume = {945},
	number = {2},
	eid = {159},
	pages = {159},
	doi = {10.3847/1538-4357/acb68c},
	archivePrefix = {arXiv},
	eprint = {2209.02675},
	primaryClass = {astro-ph.CO},
	adsurl = {https://ui.adsabs.harvard.edu/abs/2023ApJ...945..159N}
}

@ARTICLE{Fiacconi++16,
	author = {{Fiacconi}, Davide and {Madau}, Piero and {Potter}, Doug and {Stadel}, Joachim},
	title = "{Cold Dark Matter Substructures in Early-type Galaxy Halos}",
	journal = {\apj},
	year = 2016,
	month = jun,
	volume = {824},
	number = {2},
	eid = {144},
	pages = {144},
	doi = {10.3847/0004-637X/824/2/144},
	archivePrefix = {arXiv},
	eprint = {1602.03526},
	primaryClass = {astro-ph.GA},
	adsurl = {https://ui.adsabs.harvard.edu/abs/2016ApJ...824..144F}
}

@ARTICLE{Benson20,
	author = {{Benson}, Andrew J.},
	title = "{The normalization and slope of the dark matter (sub-)halo mass function on sub-galactic scales}",
	journal = {\mnras},
	year = 2020,
	month = mar,
	volume = {493},
	number = {1},
	pages = {1268-1276},
	doi = {10.1093/mnras/staa341},
	archivePrefix = {arXiv},
	eprint = {1911.04579},
	primaryClass = {astro-ph.GA},
	adsurl = {https://ui.adsabs.harvard.edu/abs/2020MNRAS.493.1268B}
}

@ARTICLE{Springel++08,
	author = {{Springel}, V. and {Wang}, J. and {Vogelsberger}, M. and {Ludlow}, A. and {Jenkins}, A. and {Helmi}, A. and {Navarro}, J.~F. and {Frenk}, C.~S. and {White}, S.~D.~M.},
	title = "{The Aquarius Project: the subhaloes of galactic haloes}",
	journal = {\mnras},
	year = 2008,
	month = dec,
	volume = {391},
	number = {4},
	pages = {1685-1711},
	doi = {10.1111/j.1365-2966.2008.14066.x},
	archivePrefix = {arXiv},
	eprint = {0809.0898},
	primaryClass = {astro-ph},
	adsurl = {https://ui.adsabs.harvard.edu/abs/2008MNRAS.391.1685S}
}

@ARTICLE{Bode++01,
	author = {{Bode}, Paul and {Ostriker}, Jeremiah P. and {Turok}, Neil},
	title = "{Halo Formation in Warm Dark Matter Models}",
	journal = {\apj},
	year = 2001,
	month = jul,
	volume = {556},
	number = {1},
	pages = {93-107},
	doi = {10.1086/321541},
	archivePrefix = {arXiv},
	eprint = {astro-ph/0010389},
	primaryClass = {astro-ph},
	adsurl = {https://ui.adsabs.harvard.edu/abs/2001ApJ...556...93B}
}

@ARTICLE{Du++24,
	author = {{Du}, Xiaolong and {Benson}, Andrew and {Zeng}, Zhichao Carton and {Treu}, Tommaso and {Peter}, Annika H.~G. and {Mace}, Charlie and {Jiang}, Fangzhou and {Yang}, Shengqi and {Gannon}, Charles and {Gilman}, Daniel and {Nierenberg}, Anna. M. and {Nadler}, Ethan O.},
	title = "{Tidal evolution of cored and cuspy dark matter halos}",
	journal = {\prd},
	year = 2024,
	month = jul,
	volume = {110},
	number = {2},
	eid = {023019},
	pages = {023019},
	doi = {10.1103/PhysRevD.110.023019},
	archivePrefix = {arXiv},
	eprint = {2403.09597},
	primaryClass = {astro-ph.GA},
	adsurl = {https://ui.adsabs.harvard.edu/abs/2024PhRvD.110b3019D}
}

@ARTICLE{Benson12,
	author = {{Benson}, Andrew J.},
	title = "{G ALACTICUS: A semi-analytic model of galaxy formation}",
	journal = {\na},
	year = 2012,
	month = feb,
	volume = {17},
	number = {2},
	pages = {175-197},
	doi = {10.1016/j.newast.2011.07.004},
	archivePrefix = {arXiv},
	eprint = {1008.1786},
	primaryClass = {astro-ph.CO},
	adsurl = {https://ui.adsabs.harvard.edu/abs/2012NewA...17..175B}
}

@ARTICLE{Gannon++25,
	author = {{Gannon}, Charles and {Nierenberg}, Anna and {Benson}, Andrew and {Keeley}, Ryan and {Du}, Xiaolong and {Gilman}, Daniel},
	title = "{Dark matter substructure: A lensing perspective}",
	journal = {\prd},
	year = 2025,
	month = jul,
	volume = {112},
	number = {2},
	eid = {023532},
	pages = {023532},
	doi = {10.1103/kk2n-q4ps},
	archivePrefix = {arXiv},
	eprint = {2501.17362},
	primaryClass = {astro-ph.GA},
	adsurl = {https://ui.adsabs.harvard.edu/abs/2025PhRvD.112b3532G}
}

@ARTICLE{Baltz++09,
	author = {{Baltz}, Edward A. and {Marshall}, Phil and {Oguri}, Masamune},
	title = "{Analytic models of plausible gravitational lens potentials}",
	journal = {\jcap},
	year = 2009,
	month = jan,
	volume = {2009},
	number = {1},
	eid = {015},
	pages = {015},
	doi = {10.1088/1475-7516/2009/01/015},
	archivePrefix = {arXiv},
	eprint = {0705.0682},
	primaryClass = {astro-ph},
	adsurl = {https://ui.adsabs.harvard.edu/abs/2009JCAP...01..015B}
}

@ARTICLE{Diemer18,
	author = {{Diemer}, Benedikt},
	title = "{COLOSSUS: A Python Toolkit for Cosmology, Large-scale Structure, and Dark Matter Halos}",
	journal = {\apjs},
	year = 2018,
	month = dec,
	volume = {239},
	number = {2},
	eid = {35},
	pages = {35},
	doi = {10.3847/1538-4365/aaee8c},
	archivePrefix = {arXiv},
	eprint = {1712.04512},
	primaryClass = {astro-ph.CO},
	adsurl = {https://ui.adsabs.harvard.edu/abs/2018ApJS..239...35D}
}

@article{astropy:2013,
	Adsurl = {http://adsabs.harvard.edu/abs/2013A%26A...558A..33A},
	Archiveprefix = {arXiv},
	Author = {{Astropy Collaboration} and {Robitaille}, T.~P. and {Tollerud}, E.~J. and {Greenfield}, P. and {Droettboom}, M. and {Bray}, E. and {Aldcroft}, T. and {Davis}, M. and {Ginsburg}, A. and {Price-Whelan}, A.~M. and {Kerzendorf}, W.~E. and {Conley}, A. and {Crighton}, N. and {Barbary}, K. and {Muna}, D. and {Ferguson}, H. and {Grollier}, F. and {Parikh}, M.~M. and {Nair}, P.~H. and {Unther}, H.~M. and {Deil}, C. and {Woillez}, J. and {Conseil}, S. and {Kramer}, R. and {Turner}, J.~E.~H. and {Singer}, L. and {Fox}, R. and {Weaver}, B.~A. and {Zabalza}, V. and {Edwards}, Z.~I. and {Azalee Bostroem}, K. and {Burke}, D.~J. and {Casey}, A.~R. and {Crawford}, S.~M. and {Dencheva}, N. and {Ely}, J. and {Jenness}, T. and {Labrie}, K. and {Lim}, P.~L. and {Pierfederici}, F. and {Pontzen}, A. and {Ptak}, A. and {Refsdal}, B. and {Servillat}, M. and {Streicher}, O.},
	Doi = {10.1051/0004-6361/201322068},
	Eid = {A33},
	Eprint = {1307.6212},
	Journal = {\aap},
	Month = oct,
	Pages = {A33},
	Primaryclass = {astro-ph.IM},
	Title = {{Astropy: A community Python package for astronomy}},
	Volume = 558,
	Year = 2013}

@ARTICLE{astropy:2018,
	author = {{Astropy Collaboration} and {Price-Whelan}, A.~M. and
	{Sip{\H{o}}cz}, B.~M. and {G{\"u}nther}, H.~M. and {Lim}, P.~L. and
	{Crawford}, S.~M. and {Conseil}, S. and {Shupe}, D.~L. and
	{Craig}, M.~W. and {Dencheva}, N. and {Ginsburg}, A. and {Vand
	erPlas}, J.~T. and {Bradley}, L.~D. and {P{\'e}rez-Su{\'a}rez}, D. and
	{de Val-Borro}, M. and {Aldcroft}, T.~L. and {Cruz}, K.~L. and
	{Robitaille}, T.~P. and {Tollerud}, E.~J. and {Ardelean}, C. and
	{Babej}, T. and {Bach}, Y.~P. and {Bachetti}, M. and {Bakanov}, A.~V. and
	{Bamford}, S.~P. and {Barentsen}, G. and {Barmby}, P. and
	{Baumbach}, A. and {Berry}, K.~L. and {Biscani}, F. and {Boquien}, M. and
	{Bostroem}, K.~A. and {Bouma}, L.~G. and {Brammer}, G.~B. and
	{Bray}, E.~M. and {Breytenbach}, H. and {Buddelmeijer}, H. and
	{Burke}, D.~J. and {Calderone}, G. and {Cano Rodr{\'\i}guez}, J.~L. and
	{Cara}, M. and {Cardoso}, J.~V.~M. and {Cheedella}, S. and {Copin}, Y. and
	{Corrales}, L. and {Crichton}, D. and {D'Avella}, D. and {Deil}, C. and
	{Depagne}, {\'E}. and {Dietrich}, J.~P. and {Donath}, A. and
	{Droettboom}, M. and {Earl}, N. and {Erben}, T. and {Fabbro}, S. and
	{Ferreira}, L.~A. and {Finethy}, T. and {Fox}, R.~T. and
	{Garrison}, L.~H. and {Gibbons}, S.~L.~J. and {Goldstein}, D.~A. and
	{Gommers}, R. and {Greco}, J.~P. and {Greenfield}, P. and
	{Groener}, A.~M. and {Grollier}, F. and {Hagen}, A. and {Hirst}, P. and
	{Homeier}, D. and {Horton}, A.~J. and {Hosseinzadeh}, G. and {Hu}, L. and
	{Hunkeler}, J.~S. and {Ivezi{\'c}}, {\v{Z}}. and {Jain}, A. and
	{Jenness}, T. and {Kanarek}, G. and {Kendrew}, S. and {Kern}, N.~S. and
	{Kerzendorf}, W.~E. and {Khvalko}, A. and {King}, J. and {Kirkby}, D. and
	{Kulkarni}, A.~M. and {Kumar}, A. and {Lee}, A. and {Lenz}, D. and
	{Littlefair}, S.~P. and {Ma}, Z. and {Macleod}, D.~M. and
	{Mastropietro}, M. and {McCully}, C. and {Montagnac}, S. and
	{Morris}, B.~M. and {Mueller}, M. and {Mumford}, S.~J. and {Muna}, D. and
	{Murphy}, N.~A. and {Nelson}, S. and {Nguyen}, G.~H. and
	{Ninan}, J.~P. and {N{\"o}the}, M. and {Ogaz}, S. and {Oh}, S. and
	{Parejko}, J.~K. and {Parley}, N. and {Pascual}, S. and {Patil}, R. and
	{Patil}, A.~A. and {Plunkett}, A.~L. and {Prochaska}, J.~X. and
	{Rastogi}, T. and {Reddy Janga}, V. and {Sabater}, J. and
	{Sakurikar}, P. and {Seifert}, M. and {Sherbert}, L.~E. and
	{Sherwood-Taylor}, H. and {Shih}, A.~Y. and {Sick}, J. and
	{Silbiger}, M.~T. and {Singanamalla}, S. and {Singer}, L.~P. and
	{Sladen}, P.~H. and {Sooley}, K.~A. and {Sornarajah}, S. and
	{Streicher}, O. and {Teuben}, P. and {Thomas}, S.~W. and
	{Tremblay}, G.~R. and {Turner}, J.~E.~H. and {Terr{\'o}n}, V. and
	{van Kerkwijk}, M.~H. and {de la Vega}, A. and {Watkins}, L.~L. and
	{Weaver}, B.~A. and {Whitmore}, J.~B. and {Woillez}, J. and
	{Zabalza}, V. and {Astropy Contributors}},
	title = "{The Astropy Project: Building an Open-science Project and Status of the v2.0 Core Package}",
	journal = {\aj},
	year = 2018,
	month = sep,
	volume = {156},
	number = {3},
	eid = {123},
	pages = {123},
	doi = {10.3847/1538-3881/aabc4f},
	archivePrefix = {arXiv},
	eprint = {1801.02634},
	primaryClass = {astro-ph.IM},
	adsurl = {https://ui.adsabs.harvard.edu/abs/2018AJ....156..123A}
}

@ARTICLE{astropy:2022,
	author = {{Astropy Collaboration} and {Price-Whelan}, Adrian M. and {Lim}, Pey Lian and {Earl}, Nicholas and {Starkman}, Nathaniel and {Bradley}, Larry and {Shupe}, David L. and {Patil}, Aarya A. and {Corrales}, Lia and {Brasseur}, C.~E. and {N{"o}the}, Maximilian and {Donath}, Axel and {Tollerud}, Erik and {Morris}, Brett M. and {Ginsburg}, Adam and {Vaher}, Eero and {Weaver}, Benjamin A. and {Tocknell}, James and {Jamieson}, William and {van Kerkwijk}, Marten H. and {Robitaille}, Thomas P. and {Merry}, Bruce and {Bachetti}, Matteo and {G{"u}nther}, H. Moritz and {Aldcroft}, Thomas L. and {Alvarado-Montes}, Jaime A. and {Archibald}, Anne M. and {B{'o}di}, Attila and {Bapat}, Shreyas and {Barentsen}, Geert and {Baz{'a}n}, Juanjo and {Biswas}, Manish and {Boquien}, M{'e}d{'e}ric and {Burke}, D.~J. and {Cara}, Daria and {Cara}, Mihai and {Conroy}, Kyle E. and {Conseil}, Simon and {Craig}, Matthew W. and {Cross}, Robert M. and {Cruz}, Kelle L. and {D'Eugenio}, Francesco and {Dencheva}, Nadia and {Devillepoix}, Hadrien A.~R. and {Dietrich}, J{"o}rg P. and {Eigenbrot}, Arthur Davis and {Erben}, Thomas and {Ferreira}, Leonardo and {Foreman-Mackey}, Daniel and {Fox}, Ryan and {Freij}, Nabil and {Garg}, Suyog and {Geda}, Robel and {Glattly}, Lauren and {Gondhalekar}, Yash and {Gordon}, Karl D. and {Grant}, David and {Greenfield}, Perry and {Groener}, Austen M. and {Guest}, Steve and {Gurovich}, Sebastian and {Handberg}, Rasmus and {Hart}, Akeem and {Hatfield-Dodds}, Zac and {Homeier}, Derek and {Hosseinzadeh}, Griffin and {Jenness}, Tim and {Jones}, Craig K. and {Joseph}, Prajwel and {Kalmbach}, J. Bryce and {Karamehmetoglu}, Emir and {Ka{l}uszy{'n}ski}, Miko{l}aj and {Kelley}, Michael S.~P. and {Kern}, Nicholas and {Kerzendorf}, Wolfgang E. and {Koch}, Eric W. and {Kulumani}, Shankar and {Lee}, Antony and {Ly}, Chun and {Ma}, Zhiyuan and {MacBride}, Conor and {Maljaars}, Jakob M. and {Muna}, Demitri and {Murphy}, N.~A. and {Norman}, Henrik and {O'Steen}, Richard and {Oman}, Kyle A. and {Pacifici}, Camilla and {Pascual}, Sergio and {Pascual-Granado}, J. and {Patil}, Rohit R. and {Perren}, Gabriel I. and {Pickering}, Timothy E. and {Rastogi}, Tanuj and {Roulston}, Benjamin R. and {Ryan}, Daniel F. and {Rykoff}, Eli S. and {Sabater}, Jose and {Sakurikar}, Parikshit and {Salgado}, Jes{'u}s and {Sanghi}, Aniket and {Saunders}, Nicholas and {Savchenko}, Volodymyr and {Schwardt}, Ludwig and {Seifert-Eckert}, Michael and {Shih}, Albert Y. and {Jain}, Anany Shrey and {Shukla}, Gyanendra and {Sick}, Jonathan and {Simpson}, Chris and {Singanamalla}, Sudheesh and {Singer}, Leo P. and {Singhal}, Jaladh and {Sinha}, Manodeep and {Sip{H{o}}cz}, Brigitta M. and {Spitler}, Lee R. and {Stansby}, David and {Streicher}, Ole and {Sumak}, Jani and {Swinbank}, John D. and {Taranu}, Dan S. and {Tewary}, Nikita and {Tremblay}, Grant R. and {Val-Borro}, Miguel de and {Van Kooten}, Samuel J. and {Vasovi{'c}}, Zlatan and {Verma}, Shresth and {de Miranda Cardoso}, Jos{'e} Vin{'i}cius and {Williams}, Peter K.~G. and {Wilson}, Tom J. and {Winkel}, Benjamin and {Wood-Vasey}, W.~M. and {Xue}, Rui and {Yoachim}, Peter and {Zhang}, Chen and {Zonca}, Andrea and {Astropy Project Contributors}},
	title = "{The Astropy Project: Sustaining and Growing a Community-oriented Open-source Project and the Latest Major Release (v5.0) of the Core Package}",
	journal = {\apj},
	year = 2022,
	month = aug,
	volume = {935},
	number = {2},
	eid = {167},
	pages = {167},
	doi = {10.3847/1538-4357/ac7c74},
	archivePrefix = {arXiv},
	eprint = {2206.14220},
	primaryClass = {astro-ph.IM},
	adsurl = {https://ui.adsabs.harvard.edu/abs/2022ApJ...935..167A}
}

\end{document}